\documentclass[amssymb,twocolumn,aps,nofootinbib]{revtex4}
\usepackage{graphics,color,graphicx,amsmath}
\usepackage{subcaption}
\usepackage{natbib}
\usepackage{verbatim}
\usepackage{graphicx}
\usepackage[above,below]{placeins}
\usepackage{float}
\usepackage{caption}
\usepackage{hyperref}

\begin{document}

\title{Not engaging with problems in the lab: Students' navigation of conflicting data and models}








\author{Anna McLean Phillips}
\email{AnnaPhillips@gmail.com}
\affiliation{Laboratory of Atomic and Solid State Physics, Cornell University}

\author{Meagan Sundstrom}
\affiliation{Laboratory of Atomic and Solid State Physics, Cornell University}

\author{David G. Wu}
\affiliation{Laboratory of Atomic and Solid State Physics, Cornell University}

\author{N. G. Holmes}
\affiliation{Laboratory of Atomic and Solid State Physics, Cornell University}

\date{\today}

\begin{abstract}
With the adoption of instructional laboratories (labs) that require students to make their own decisions, there is a need to better understand students' activities as they make sense of their data and decide how to proceed. In particular, understanding when students do not engage productively with unexpected data may provide insights into how to better support students in more open-ended labs. 
We examine video and audio data from groups within a lab session where students were expected to find data inconsistent with the predictions of two models. In prior work, we examined the actions of the four groups that productively grapple with this designed problem \cite{ProblematizingPERC}. Here, we analyze the engagement of the three groups that do not. We conducted three phases of analysis: 1) documenting large scale behaviors and time spent in on-topic discussion, 2) analyzing interactions with the teaching assistant, and 3) identifying students' framing--their expectations for what is taking place--when they were discussing their data.  
Our Phase 1 and 2 analysis show only minor differences between the groups that engaged with the problem and those that did not. Our Phase 3 analysis demonstrated that the groups that did not engage with the problem framed the lab activity as about confirming a known result or as a series of hoops to jump through to fulfill assignment requirements. Implications for instruction include supporting teaching assistants to attend to students' framing and agency within laboratory classrooms.
\end{abstract}

\maketitle

\section{Introduction}

Institutions are presently reforming their instructional laboratories (labs) to shift away from ``cookbook-style" activities \cite{holmes2018introductory, clough2002using}, where students follow a set procedure to obtain an expected result, to activities that require students to make their own experimental decisions and conclusions. With these reforms underway, there is a need for additional research on how students engage in these reformed labs. Previous work has shown that when students are empowered to make their own decisions in lab, the outcomes may vary. Students can engage productively in scientific practices \cite{Brewe2008, meyer2017student, Zwickl2015, Dounas-Frazer2016, HolmesPNAS} and develop inquiry skills \cite{Bartlett2019, Etkina2006, Etkina2008}.  
In contrast, some studies have found many students engage in questionable research practices to confirm canonical phenomena, particularly when the lab activity is set up for students to identify a limitation of the canonical phenomenon~\cite{SmithStienHolmes2018, SteinPERC2018, Smith2020QRP}.

When confronted with a result that surprisingly disagrees with their expectations, students may focus on confirmation rather than engaging in what we refer to as \emph{problematizing}, the process of grappling with uncertainty and refining what they do and do not know \cite{phillips2017problematizing}. Problematizing is a core part of what it means to engage in science \cite{buck2014sure, phillips2018beyond} and in physics in particular \cite{phillips2017problematizing}. Labs provide a natural setting for students to engage in problematizing, as inconsistencies between students' expectations and their data may arise naturally or be designed into the experimental setup \cite{HolmesPNAS, Bartlett2019, Vonk2017, michelini2018labs,hardy2020data}. Here, we focus on a lab activity designed to have students uncover an inconsistency between their data and expected models in order to understand when students engage productively in problematizing. Understanding when students engage productively and when they do not is an important step in developing curricular and professional development tools to improve such labs.


In this paper, we seek to understand what takes place when we design a lab exercise in which students' data ought to conflict with the models they expect. In previous work \cite{ProblematizingPERC}, we have examined the characteristics of groups that recognize and engage meaningfully with the intended problem. Here, we look more closely at the groups that do not, with the aim of understanding how to better support students' problematizing.

To do so, we present a framework for understanding students' problematizing, followed by a presentation of several approaches we used to understand coarse- and fine-grained differences in how the groups engaged with the activity. We discuss patterns across the groups that did not problematize and conclude with implications for instruction. The behavior of groups that did not problematize demonstrates that their \emph{framing} — their expectations of what is taking place \cite{Tannen,ScherrHammer}--is to either confirm given models or jump through hoops to complete the assignment quickly. The non-problematizing groups were also never observed as perceiving or acting upon opportunities to build knowledge, an aspect of \emph{epistemic agency} \cite{stroupe2014examining,miller2018addressing}.

\subsection{Theoretical Background}

Researchers in science education have paid increasing attention to the important role of students' engagement with uncertainty and ambiguity in science classrooms as a part of authentic science practice \cite{manz2015resistance,manz2018supporting,chen2019managing,stroupe2018fostering,watkins2018positioning}. In physics in particular, the process of articulating and refining uncertainties into clear questions or problems, which we refer to as problematizing, is an important aspect of professional physics that is also present in students' engagement in science \cite{phillips2017problematizing}. Research at the K-12 level has shown even young students are capable of grappling with uncertainty with qualitative observations, quantitative data, and models \cite[e.g.,][]{metz2008narrowing, metz2004children, engle2002guiding, engle2012productive,buck2014sure,chen2020dialogic, chen2019managing}. Research on how college students engage with uncertainty is more sparse \cite{gouvea2020argumentation,phillips2017problematizing,ehrlich2018eureka} but further supports that students at all levels can \textit{productively} engage with ambiguity. 

Productivity can be framed in a variety of ways. 
In the views of Engle \& Connant, productivity is tied to where students' inquiry leads them after developing a question and is intrinsically discipline-specific \cite{engle2012productive, engle2002guiding}. Kapur \cite{kapur2008productive,kapur2016examining} argues that productivity depends on how well students' activities set them up for future learning. Manz describes ``productive uncertainty" by arguing that ``by grappling with some of the decisions scientists must make, students would make progress on scientific practices and content understandings" \citep[p. 1-2]{manz2018designing}. We build upon and expand Manz's view, taking the stance that problematizing is itself a scientific practice, though it may also involve other scientific practices, such as making arguments from evidence \cite{phillips2017problematizing,phillips2018beyond}. In order for problematizing to take place, students need only to make some progress in narrowing down what it is they are uncertain about. We consider problematizing itself to be productive, even if students do not solve the problem. Problematizing may involve building further content understanding, though it does not necessarily require it. 

Problematizing and engagement with uncertainty more broadly is often challenging for students and instructors alike, and requires active management of that uncertainty. Students and teachers may  resist the presence of ambiguity in instruction \cite{engle2012productive, manz2018supporting, chen2020dialogic}. While articulating their uncertainty can be productive for students' engagement \cite{radinsky2008students, conlin2012building, watkins2018positioning}, such expressions can carry with them social risk \cite{watkins2018positioning}. To mitigate those risks, students may ``epistemically distance" themselves from uncertain claims to place distance between themselves and their claims \cite{conlin2018making}, use humor to to deflect from their uncertainty \cite{conlin2012building} as a part of their productive engagement, or leave statements of uncertainty intentionally ambiguous or implicit \cite{buck2014sure}. Productively managing uncertainty requires attending to raising, maintaining, and reducing uncertainty at appropriate times \cite{gouvea2020argumentation, chen2019managing, chen2020dialogic}. Instructors and students can both make these uncertainty-managing moves \cite{watkins2018positioning, chen2020dialogic}. Instructors may need support and professional development to do so \cite{manz2018supporting,manz2018designing, ko2019opening, chen2020dialogic} and students may require support from their instructor \cite{watkins2018positioning, chen2020dialogic}.

In order for students to problematize, they must have authority over whether or not the ideas and information in front of them are coherent, which is a form of \emph{epistemic agency} \cite{keifert2018epistemic}. Epistemic agency refers to having authority over what counts as knowledge and how that knowledge is constructed \cite{stroupe2014examining}. Epistemic agency can be conceptualized as students acting as individual ``epistemic agents" \cite{stroupe2014examining} who generate knowledge. We can also understand epistemic agency in a collective sense  as ``students being positioned with, perceiving, and acting on, opportunities to shape the knowledge building work in their classroom community" \citep[p. 1058]{miller2018addressing} which emerges from  ``collaborative activities aimed at the creation of shared knowledge objects" \citep[p. 146]{damcsa2010shared}. That is, epistemic agency does not just lie with individuals acting as agents, but as a shared understanding that authority over what counts as knowledge rests with the group, not an instructor or other external source of knowledge. Supporting epistemic agency within K-12 classrooms has been discussed extensively within the science education research community \cite[e.g.,][]{damcsa2010shared,stroupe2014examining,miller2018addressing,stroupe2018fostering,ko2019opening}. As with the literature that foregrounds students' uncertainty, these researchers highlight that supporting agency requires responsive teaching that is challenging, though feasible, for instructors \cite{stroupe2018fostering,ko2019opening}. 

In introductory physics labs, students may be positioned to develop their own claims, models, and conclusions. In the labs we examine here, students are explicitly evaluated on whether or not their findings are justified based on their data and they are expected to develop new models to explain their data. Using the language from Miller and colleagues \cite{miller2018addressing}, students are \emph{positioned with} opportunities to build knowledge. However, as we will show, not all students \emph{perceive} or \emph{act on} those opportunities in the lab.

Students perceptions and actions are deeply tied to their framing: How they tacitly answer ``What is taking place here?" \cite{hutchison2010attending}. Framing \cite{Tannen,TannenWallat,Goffman} and epistemological framing \cite{hutchison2010attending, ScherrHammer} allow us to understand how students engage in their activities. Framing refers to how people understand what is taking place in a given situation \cite{Tannen,Goffman}. Evidence for how someone frames a situation can be drawn from their speech and behaviors \cite{Tannen,Goffman}. Epistemological framing refers to how people understand what is taking place with respect to knowledge \cite{hutchison2010attending,ScherrHammer}. Epistemological framing is closely tied to epistemic agency: if a group of students expect to reproduce known results in the lab, they regard knowledge as something that will be given to them by the instructor or written resources rather than constructed by them \cite{SmithStienHolmes2018,Smith2020, Zwickl2015}. Thus, they would be likely to perceive or act upon opportunities for epistemic agency. If students do not activate a frame for understanding lab as a knowledge-building opportunity, they may not perceive or act on opportunities that instructors or curriculum designers intend. If a student makes a bid to investigate a surprising result, the group may shift into an epistemological frame of understanding that they are investigating a new phenomenon without anticipating a given result. They would then be acting with epistemic agency. In this way, using methods drawn from literature on framing, we can describe the role of epistemic agency within groups and gain insight into how students engage with uncertainty, whether or not they do so productively. We combine this approach with quantitative analysis of students' speech and actions to identify coarse features of their engagement.

\subsection{Current Project}

In prior work \cite{ProblematizingPERC}, we examined the activities performed by lab groups who successfully took up and engaged with the uncertainty that the lab curriculum intentionally proposed.  We transcribed and coded video episodes of how students dealt with this explicit inconsistency and determined their subsequent problematizing behaviors. Among the four groups who successfully engaged with the problem, students engaged in similar activities, though in different lengths and sequences. Some behaviors we found productive lab groups to engage in when problematizing include physical reasoning, proposing a new experiment, checking experimental calculations, and consulting an external reference. These behaviors outline what students do when they engage with the uncertainty specifically put forth by the lab material; however, this analysis did not capture why the remaining three lab groups did not problematize around this uncertainty.

\begin{figure*}[t]
    \centering
    \includegraphics[width=6in]{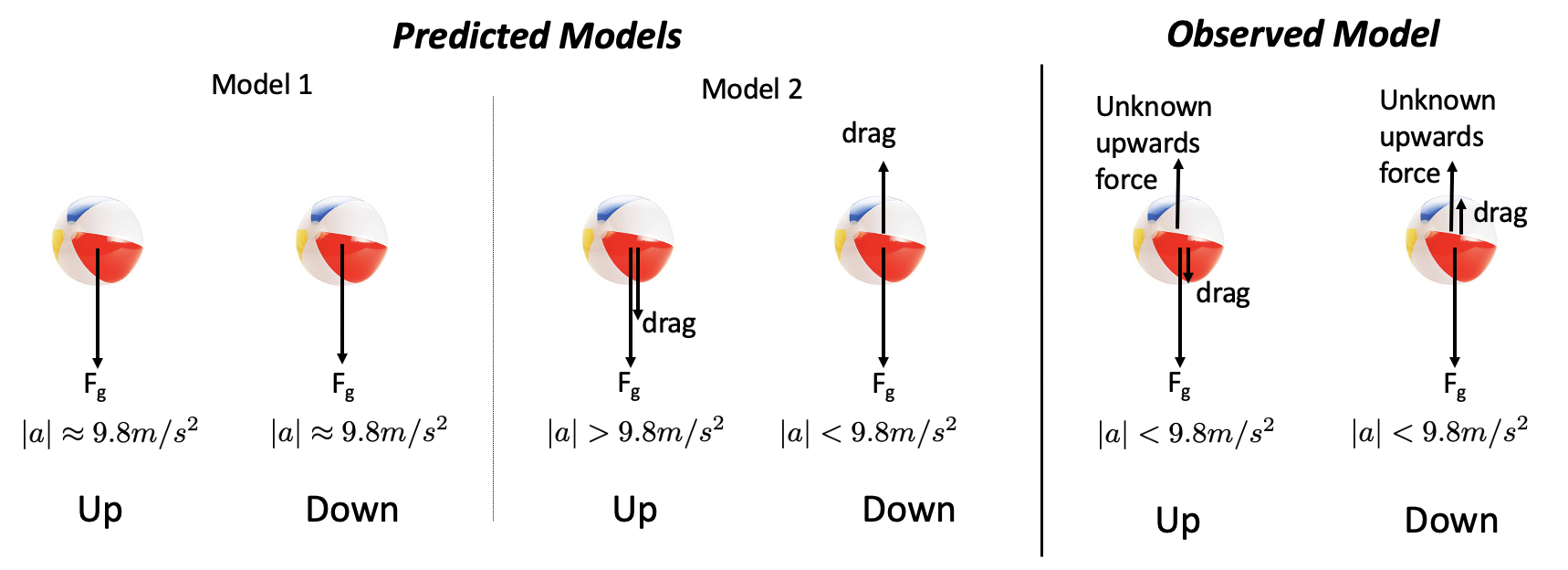}
    \caption{The students are asked to predict the accelerations of the balls using two models. Model 1 is if the ball only experiences a force of gravity. Model 2 is if the ball experiences both gravity and drag. The prediction of Model 1 is that the acceleration of the ball is a constant $9.8m/s^2$. The prediction for Model 2 is that the acceleration of the ball is greater than $9.8m/s^2$ on the way up and less than $9.8m/s^2$ on the way down. Students find that the measured acceleration is below $9.8m/s^2$ in magnitude in both directions, which indicates the presence of another upwards force.}
    \label{fig:Models}
\end{figure*}

Having developed some understanding of the dynamics of problematizing in the lab, we turn now to understand the groups that did not engage in this practice. To borrow a phrase, all of the problematizing groups were somewhat alike; every non-problematizing group initially appeared to avoid the problem in their own way. Here, we focus on these three unproductive groups to answer the following questions: \textbf{How can we characterize their engagement? What may have contributed to their lack of problematizing around the designed problem?}

To answer these questions, we analyzed the video data in three phases: 1) analyzing large scale behaviors throughout the lab session, 2) analyzing student interactions with the teaching assistant (TA), and 3) analyzing the framing of students during discussions of the data that would allow them to identify the designed problem.

In line with our phases of analysis, we present our methods and findings from each phase in this order. We then summarize our findings for each of the three groups that did not problematize. From Phase 1, we identified that groups that did not problematize exhibited differences in the amount of speech, the timing of data collection and discussion, and amount of on topic conversation. In Phase 2, we examined the students' interactions with the TA, finding no clear pattern amongst the problematizing and non-problematizing groups. In Phase 3, we determined that the groups that did not problematize were primarily in frames that reflected goals of confirming expected models \cite{SmithPERC2020,SmithStienHolmes2018, Smith2020QRP} or jumping through hoops to complete the activity as quickly as possible.

\section{Context \& Data Collection}

The data for this study come from the laboratory component of an introductory mechanics course for engineers at Cornell University. There were 19 students in the section on this day, spread across seven groups of two to three students. This section had eight female students and eleven male students. The TA teaching this section was an experienced teacher familiar with physics education research and the lecture portion of the course was taught as an interactive lecture. 

The lab activities in this course are designed to promote laboratory skills development and critical thinking \cite{holmes2019operationalizing} and require students to design and execute their own experiments. This particular session was the fourth in the semester and was the first session of the second major unit of the lab curriculum. At the start of the lab session, the TA told students to form groups with students with whom they had not previously worked.

In this lab, students explored the free-fall acceleration of different objects in flight and were asked to evaluate two physical models: one in which gravity is the only force acting on the object and another in which both gravity and the force of air drag act on the object.\footnote{This activity is based on one originally developed by David Marasco.} Students made predictions of what they expect their data to show, namely they determined whether each model predicts the free-fall acceleration of the object to be greater than, less than, or equal to the magnitude of gravitational acceleration, \emph{g}, when the ball is traveling both upward and downward. Force diagrams similar to those that students drew to inform their predictions are shown in Fig.~\ref{fig:Models}. Students collected acceleration data for different falling objects, including basketballs and beach balls. Beach balls were made available to students because the buoyant force has a substantial impact on their motion, yielding accelerations both up and down that are less than the acceleration due to gravity, inconsistent with predictions made by both models. Sample data reflecting this effect are shown in the supplement (Sec. \ref{Supplement} A). Thus, the lab was specifically designed to encourage students to problematize: they are expected to grapple with the inconsistency between their data and the two provided models. 

We video recorded the lab session with two wall mounted cameras. In order to capture each group individually, audio recorders were placed with each of the seven groups of students. The TA also wore a portable microphone, resulting in a total of eight audio tracks. Each lab table audio track was paired with the video that best showed that group.

These recordings allowed us to see students' physical movements for all lab tables; however, we were unable to resolve facial expressions for groups far from the cameras. Clear audio recordings in group work settings are challenging due to the volume of background noise and cross talk. We were able to understand over 80\% of the speech for six of the seven groups, and over 90\% for four of the groups. This high percentage of audible talk allowed us to follow the conversations of six of the seven groups. We also had access to the lab notes for each group, which we used to understand analysis students completed if it was unclear from their verbal speech.  

We now discuss our phases of analysis (large scale behaviors, interactions with the TA, and framing), along with the findings from each phase. 

\section{Phase 1: Large scale behaviors}


\subsection{Phase 1 Methods}
\begin{figure}
    \centering
    \includegraphics[width=3in]{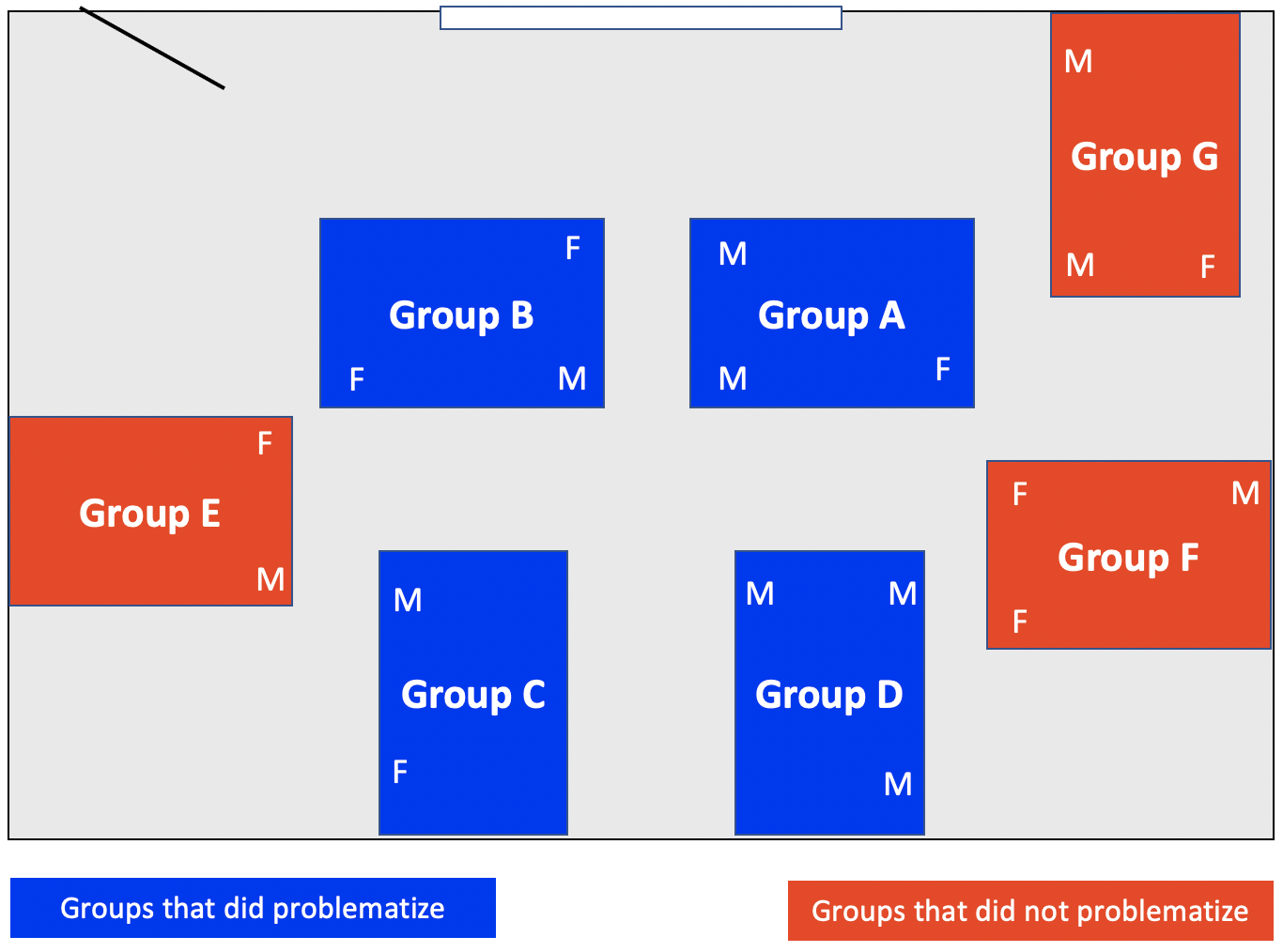}
    \caption{Diagram of the classroom layout. Numbers and approximate seating arrangement of female (F) and male (M) students are shown. Groups in blue (A, B, C, and D) did problematize around the intended problem. Groups in red (E, F, and G) did not. }
    \label{fig:Layout}
\end{figure}

One researcher watched for the entire two hour long lab session and the student activities were summarized at 5-minute intervals. From these observations, she noted when students collected and discussed their data, when students discussed the measured accelerations of the beach ball, and if they compared this result to their predictions. This led to the identification of the episodes analyzed in Ref \cite{ProblematizingPERC} and the moments analyzed in Phase 3 (described in Sec. V). The four groups discussed in \cite{ProblematizingPERC} are referred to as Groups A, B, C, and D in the following discussion. The groups omitted from that analysis (Groups E, F, and G) did not problematize around the intended problem and are the focus of this paper. The groups, along with their configuration in the room and gender composition, are shown in Fig. \ref{fig:Layout}. While Fig. \ref{fig:Layout} shows the non-problematizing groups were all along the edge of the room, we do not believe this was a substantial factor in students' problematizing as the TA's interactions with students do not appear to depend on classroom layout, as discussed in Section \ref{sec:TA}. Separate analysis of this video in other research has shown the number and duration of interactions between groups also does not depend on classroom layout \cite{Networks}.

To more closely examine students' engagement in the laboratory activity, we watched each full video again, beginning when the class finished a full group discussion about a preliminary activity on confirmation bias. This was 40 minutes into the two hour session. As most groups did not finish on time, this left between 1 hour and 22 minutes to 1 hour and 32 minutes of video. Using the video coding software BORIS \cite{friard2016boris}, we coded for audible and inaudible speech, silence (longer than 5 seconds), on- and off-topic talk (if possible), moments of data collection (inclusive of troubleshooting with the apparatus), and if the TA, lab manager, or a student from another group was at the table. A 10-minute sample of data containing each type of event was coded by a second coder to test reliability. We used the BORIS built-in tool for calculating Cohen's $\kappa$, which samples the data and checks the codes at those times. Various sample rates between every 0.5 seconds (1200 comparison points) and 5 seconds (120 comparison points) all yielded $\kappa > 0.849$. Other measures of interrater reliability are provided in the supplement (Sec. \ref{SupplementCoding}).

To complement this analysis, we performed a word count of each group's lab notes to determine if differences in apparent time on task were reflected in their written work. In this way, we were able to identify large scale differences between the engagement of the different groups. Timelines showing this full coding for each group can be found in the supplement (Sec. \ref{SupplementGroups}). 

\subsection{Phase 1 Findings}

\subsubsection{Speech and silence}

\begin{figure}[h]
    \centering
    \includegraphics[width=3.5in]{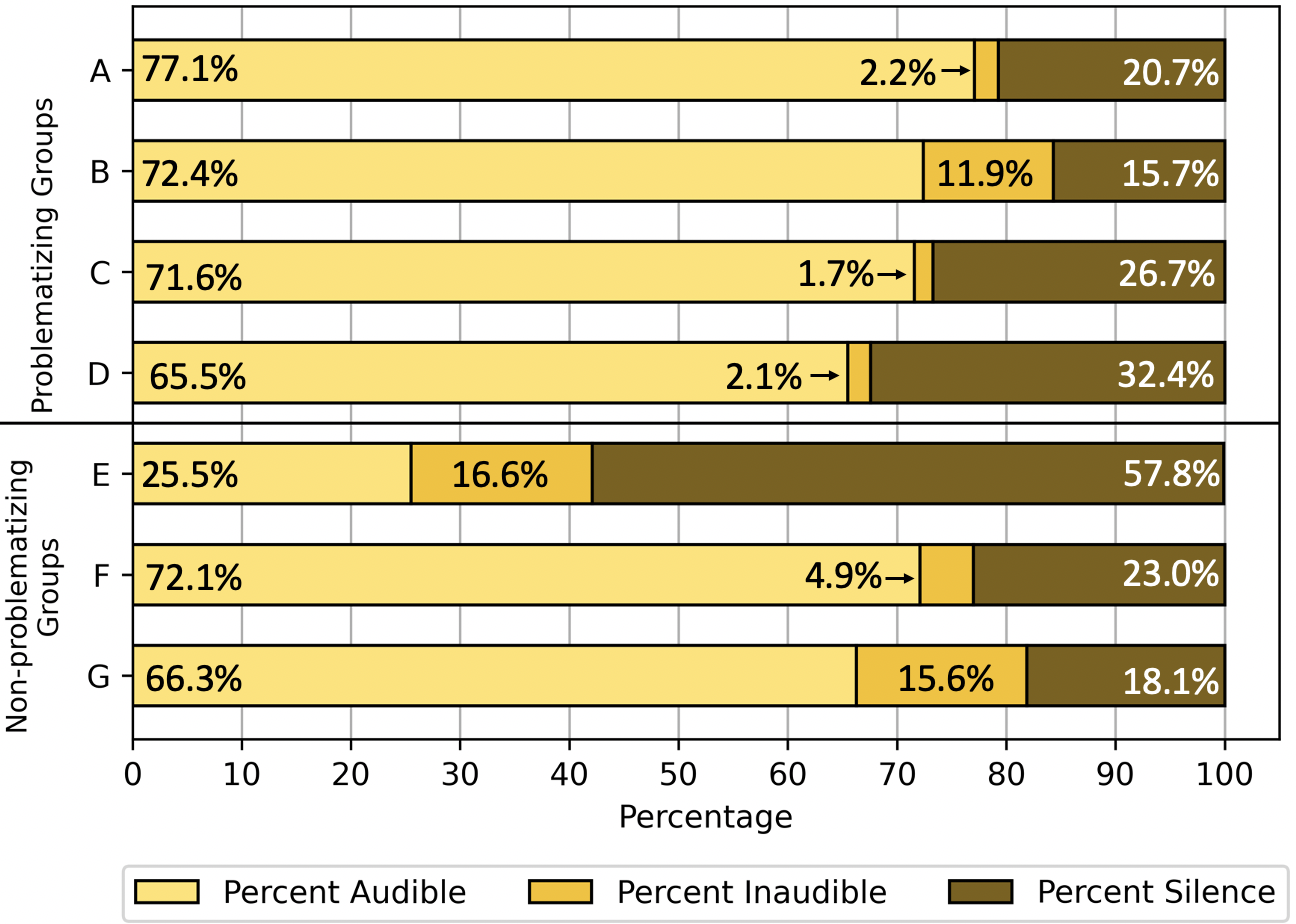}
    \caption{Percent of time spent working in groups that was audible and inaudible speech and silence.  Group E spends the majority of their time in silence, while all other groups spend the majority of their time in discussions audible on recordings.}
    \label{fig:Audible}
\end{figure}

\begin{figure}[h]
    \centering
    \includegraphics[width=3in]{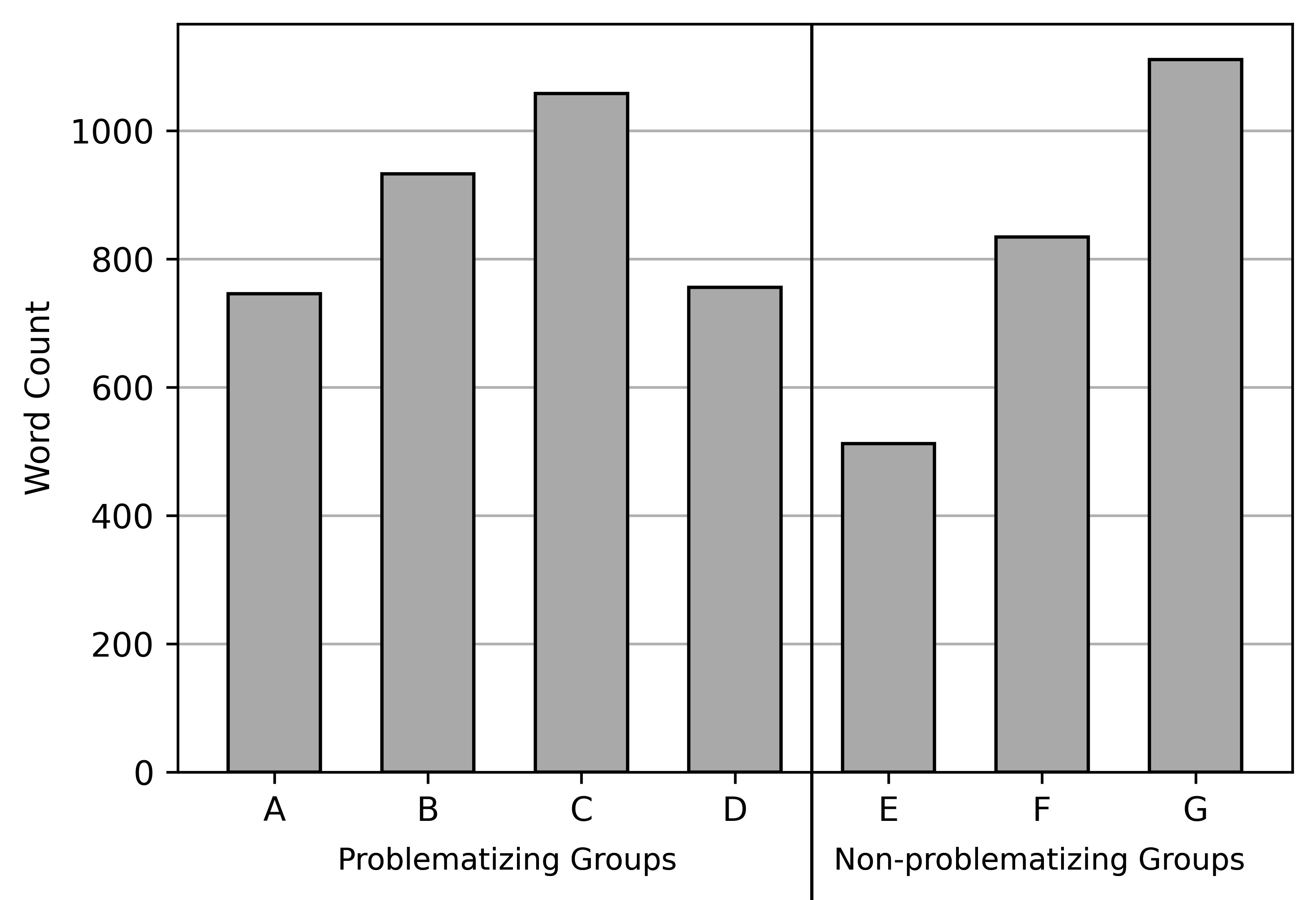}
    \caption{Word count of text in lab notes for each group. Equations and data tables are excluded from the counts. Group E's lab notes contained the least text of any set of notes.}
    \label{fig:Words}
\end{figure}

Six of the seven tables spent the majority of their time in audible discussion. The exception is Group E, whose audio quality is often poor.
While this audio quality prevented us from determining if their conversations were on or off topic, we did observe that they spent over 50\% of the their time in the lab in silence, as shown in Fig. \ref{fig:Audible}. Of the time we could identify that the students were speaking, approximately 40\% is inaudible. Much of what is audible is spent with the TA, and the silence and inaudible speech are spread throughout the lab session (a full timeline of Group E's speech can be found in the Supplementary Material, Fig. \ref{fig:GroupF}). 
The extended silence suggests this group was not as engaged in the lab activity as the other tables. That was further supported by their lab notes: their notes were shorter than any other group, as shown in Fig. \ref{fig:Words}. To assist the reader in connecting our findings for the non-problematizing groups across our phases of analysis, we refer to Group E as the \emph{Quiet} Group E from here on.


\subsubsection{On- and off-topic conversations}
\begin{figure}[ht]
\centering
    \includegraphics[width=3.5in]{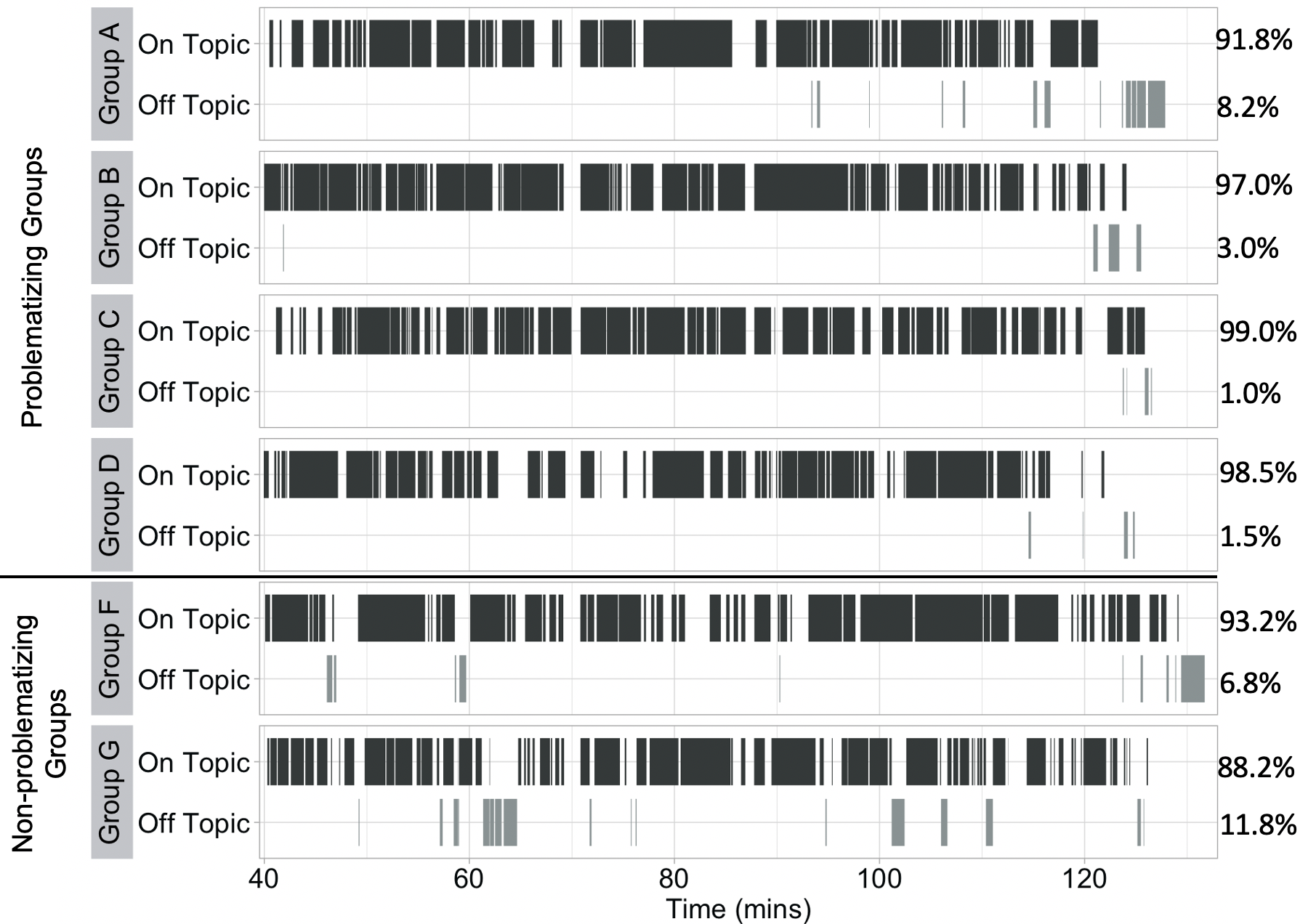}
    \caption{Timeline of on- and off-topic speech for the six codeable groups, along with the cumulative percentage of speech that is on- or off-topic. Silence and extended inaudible speech are excluded from the percentages and appear as gaps in the timeline. Groups F and G are the only two to spend substantial time off topic in the middle of the lab. Off-topic discussion at the end of the lab for all groups includes discussion about the learning management software used for submitting their reports.}
    \label{fig:Topic}
\end{figure}

In all code-able groups, the vast majority of their speech is on topic, as seen in Fig. \ref{fig:Topic}. Of the four groups that problematized around the intended inconsistency, three spent 3\% or less of their time in off-topic conversation. Group A, who spent 8\% of their time off topic, spends a longer time at the end of the lab session cleaning up their table than other groups. The bulk of their their off-topic speech was during this time and therefore did not impact their engagement with the lab activity. The timing of off-topic speech for all groups is shown in figures in the supplement (Sec. \ref{SupplementGroups}).

Group G
did not problematize and has the most time off topic of any group. In addition
, the nature and timing of their off-topic conversations differ from those of the other groups. Many other groups' off-topic conversations are somewhat related to the lab at hand or are otherwise simple small talk. For example, in Group D, one student asks another where he got his calculator. Students in Group A briefly discuss napping and getting coffee. All of the audible groups spend some time discussing their dislike of the learning management system at the end of the lab session. In Group F, who spends the second most time off topic, two students veer from discussing the basketballs and beach balls to discussing their shared experiences playing tennis. While the third student does not participate in this discussion, he does not express any discomfort. As with Group F, two students in 
Group G have off topic conversations that exclude the third member. However, these conversations are centered on alcohol and their plans for spring break and, unlike the off-topic conversation about tennis in Group F, there are multiple conversations on these topics. In one of these discussions, the female student declares, ``I feel like I'm in a frat," indicating potential discomfort. These conversations take place out of earshot of the TA. We refer to Group G as the \textit{Social} Group from now on, due to their sporadic off-topic conversations throughout the lab. We refer to Group F as the \emph{Uncertain Group}, for reasons that will be discussed in Section \ref{sec:OtherBehaviors}.

The timing of the off-topic conversation also differs between the problematizing and non-problematizing groups. Off-topic conversations in most groups are concentrated at the end of the lab when students are submitting their lab reports and cleaning up, as shown in Fig. \ref{fig:Topic} (This is when most complaints about the learning management software take place, as it takes some time for the students to follow the steps to submit their notes). The exceptions to this are often brief, such as Group A's comments about napping and Group D's conversation about calculators. Group F and Group G are the only tables to have off-topic conversations exceeding one minute in the middle of the lab period. In both groups, the off-topic conversations involved the group members of the same gender, while largely excluding the third student. 

\vspace{1pc}

\subsubsection{Data Collection}
\begin{figure}[ht]
\centering
    \includegraphics[width=3.5in]{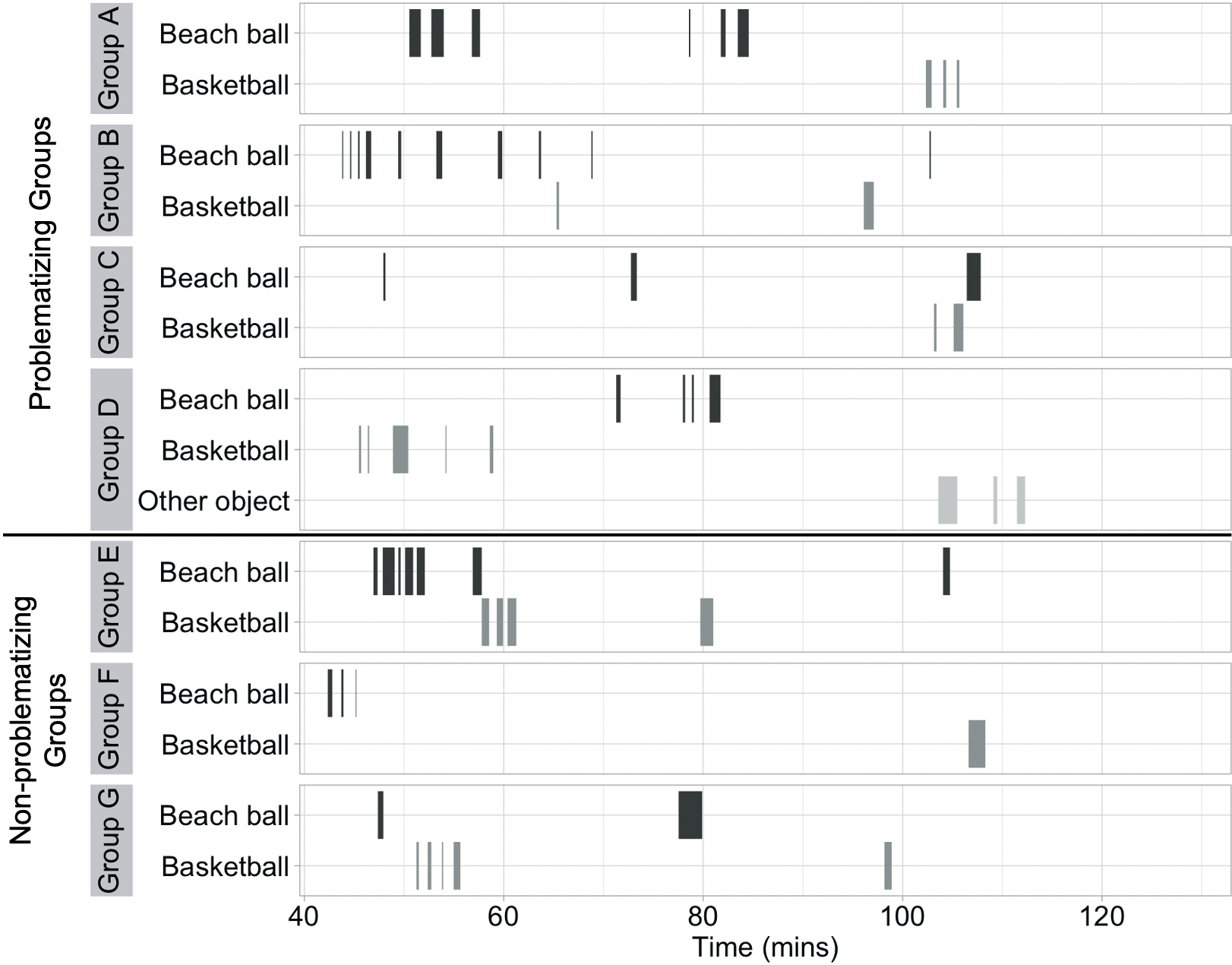}
    \caption{Timing of data collection for all groups. For most groups, data collection occurs frequently. Group F experiences a large gap in time between their collection of data for the two balls. Group G does not obtain usable data during their first attempt with the beach ball, only obtaining beach ball data to analyze beginning at 75 minutes into the lab. Group D, which does problematize, was the only group to attempt to collect data using other objects to test their ideas.}
    \label{fig:Data}
\end{figure}

Most groups collected data throughout the lab session, either to collect additional data to analyze or to confirm that their initial data set was replicable (Fig. \ref{fig:Data}). The TA discouraged this by letting groups know when they had collected enough data to conduct analysis in order to ensure students completed their work in the allotted time. The Uncertain Group F, who does not problematize, is the only group to have followed the TA's suggestion, collecting all of their data with the beach ball early in the lab and all of their data with the basketball in one round much later. They also collect the smallest amount of data relative to other groups. The Quiet Group E did not obtain usable data until over 40 minutes into the lab activity, yet did not ask for help that we can identify. Once they did obtain usable data, they did not conduct multiple trials of other balls. The fact that most groups ignored the TA's instructions is evidence that these groups took up agency that the Quiet Group E and the Uncertain Group F did not. The Social Group G's first attempt at collecting beach ball data failed and they obtained their first usable data from the beach ball much later than other groups. Unlike Groups E and F, they took multiple trials of the basketball that they used for analysis.


\subsubsection{Other behaviors}
\label{sec:OtherBehaviors}

We noted several other behaviors at this stage among the non-problematizing groups.  In the Quiet Group E, the students spend most of their time sitting on opposite sides of their lab table looking individually at their computers while in silence. In the Social Group G, one student, Brett,\footnote{All names are pseudonyms} frequently left the table to walk around room to talk to students in the other groups. This left two students, Rachel and Mike, to complete much of the lab without him and Rachel was frequently in the role of asking her male peers to return to the laboratory tasks. In Group F, the students frequently stopped their discussion to wait for assistance from the TA. This is why we refer to them as the Uncertain Group F. We hypothesized that these differences in help seeking behavior from the Group F, which indicate a reluctance to act with agency, were related to their lack of problematizing. This motivated our closer analysis of groups' interactions with the TA. 

\section{Phase 2: TA Interactions Analysis}
\label{sec:TA}

\subsection{Phase 2 Methods}

To further investigate why certain lab groups did not engage with the uncertainties built into the lab in the intended manner, we analyzed all interactions between the groups and the TA. We examined the content and duration of these interactions with the video data of the full class and the audio from a microphone worn by the TA. Using BORIS, we coded interactions between the TA and each group, recording the duration and whether the TA or the group initiated these interactions. 

In order to understand how these interactions may have impacted students' problematizing, we also coded for how long students and the TA discussed the interpretation of their results, which we defined as, ``Students and TA discuss a result by interpreting its meaning such as discussions about why a result confirms/disconfirms a model, what a statistic indicates about a dataset, or whether the result makes sense." We coded these episodes at the level of communicative events (``a series of turns in the conversation where participants, participant structure (class, group or dyad), purpose, task, orientation and/or general topic remain constant" \citep[p. 18]{hennessy2016developing}), as we required that the TA and at least one student engage in back and forth speech on a single topic. Further details about this coding can be found in the supplement (Sec. \ref{Supplement} B). An interrater reliability test on 10 minutes of data showed 100\% agreement on all codes, with start and stop times of codes differing by less than 5 seconds. Because of the full agreement, we did not calculate statistical measures of agreement.



\subsection{Phase 2 Findings}

\begin{figure}[h]
\centering
    \includegraphics[width=3.3in]{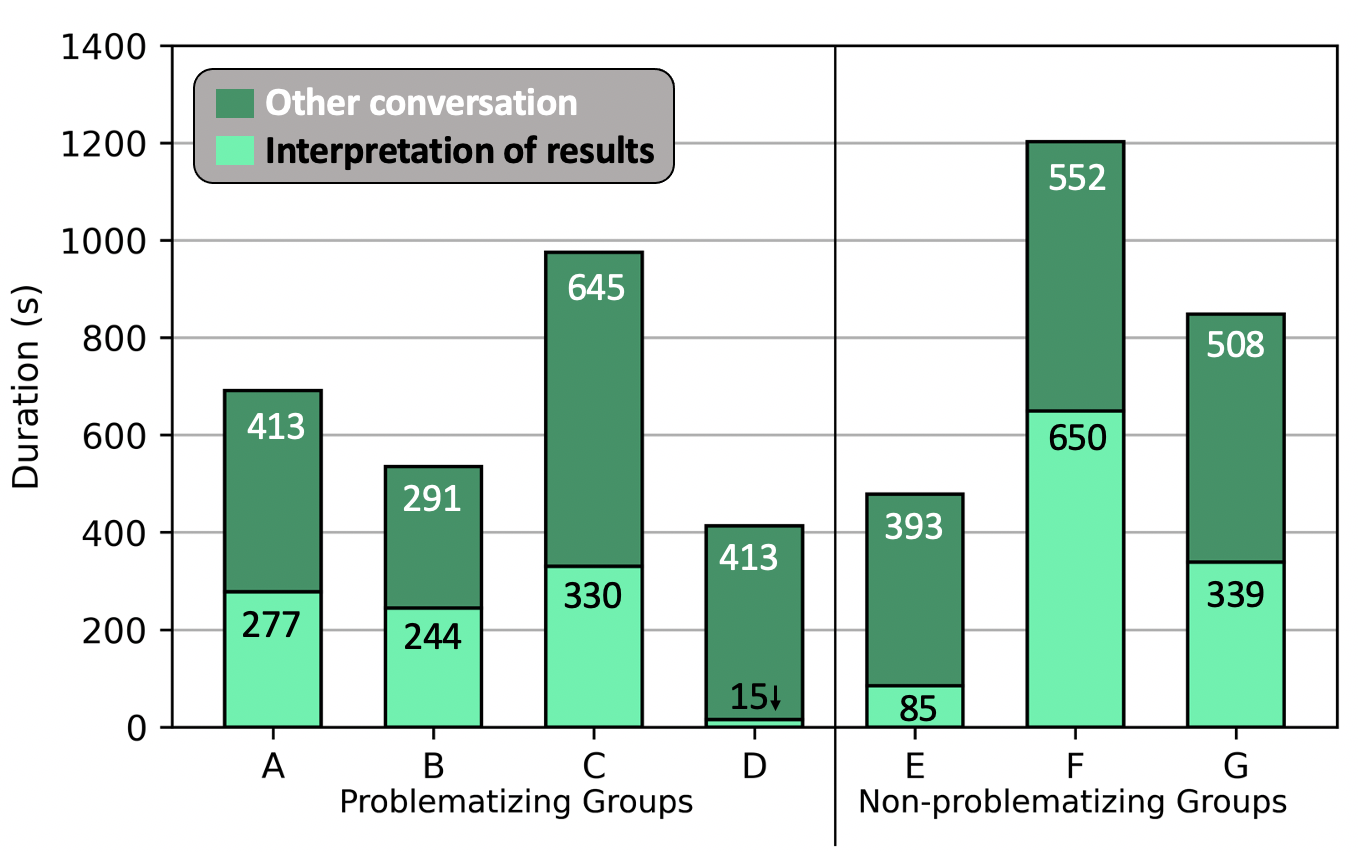}
    \caption{The time spent in other conversation is similar across all groups, with the time spent discussing interpretation of results varying from a low of 15 seconds (Group D) to a high of 650 seconds (Group F). }
    \label{fig:TAInterp}
\end{figure}

\begin{figure}[h]
\centering
    \includegraphics[width=3.3in]{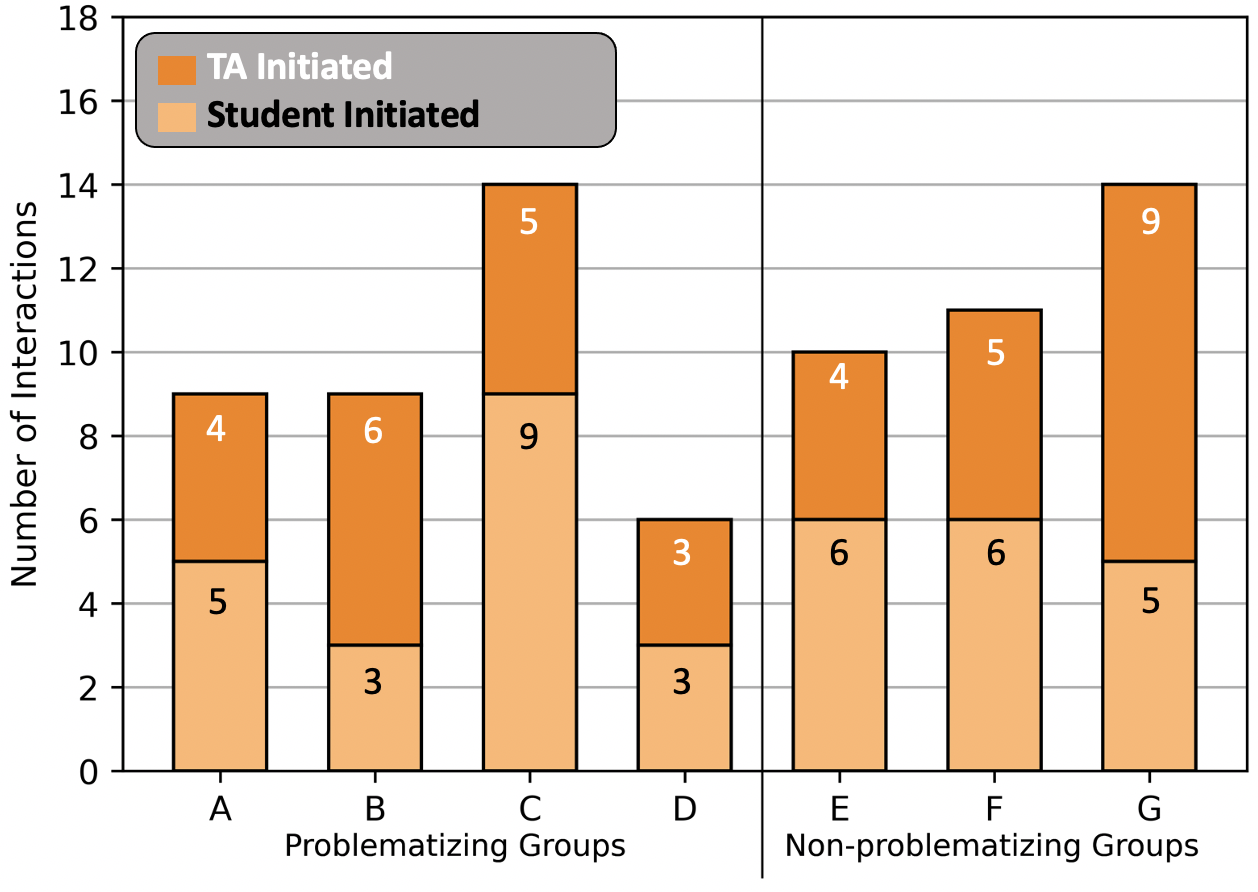}
    \caption{The number of interactions with the TA is lowest for Group D, which did problematize, and highest for Groups C, which did problematize, and G that did not.}
    \label{fig:TA}
\end{figure}

Overall, there was no clear pattern in either the number of interactions or the time or content of those discussions among groups that did and did not problematize. As seen in Fig. \ref{fig:TAInterp}, the TA spent the least amount of total time and the least amount of time discussing interpretation of results with Group D, who did problematize. The Quiet Group E, who did not problematize, spent the second least amount of time with the TA and also the second least amount of amount of time discussing interpretation of results. However, the total number of Group E's conversations and who appeared to initiate those conversations were similar to those of the other groups, as seen in Fig. \ref{fig:TA}. 

The large scale behavior analysis (Phase 1) suggested that the Uncertain Group F interacted more with the TA than the other tables and engaged in more help-seeking behaviors. This was partly confirmed in the Phase 2 analysis: the TA spent more time in conversation with Group F than the other groups and much of that additional time was spent discussing interpretation of results, as seen in Fig. \ref{fig:TAInterp}. The Uncertain Group F's help-seeking behaviors are discussed more in Phase 3 findings.

With few obvious patterns across the groups that did and did not problematize, we turned to more fine grained qualitative analysis, examining student framing during moments where they discussed their data. 
  
\section{Phase 3: Framing analysis}
\subsection{Phase 3 Methods} 

As a lens to further understand why some lab groups did not engage productively with presented inconsistencies, we examined how these groups framed the lab task. That is, we sought to answer the question: ``How do these students understand what is taking place and what are their expectations for what should happen in the lab?" 

We transcribed moments where students compared their beach ball data to the given models. For the problematizing groups (Groups A, B, C, and D), this corresponded to the episodes discussed in \cite{ProblematizingPERC}. For the Quiet Group E, no such moment was identified because of the high portion of speech that was inaudible. Instead, we provide some analysis of their discussion of basketball data with the TA. For the Uncertain Group F, we transcribed the majority of the episode, as they spend an extended period of time discussing their data but not comparing it to the models. For the Social Group G, we transcribed several shorter moments as their discussions of their beach ball data was very brief. 

To determine student framing, we examined students’ discourse and behaviors \cite{Tannen,TannenWallat,ScherrHammer} when they discussed their beach ball data. We identified three broad frames that capture differences in the groups' approaches to the lab.  As described in Ref. \cite{SmithStienHolmes2018, Smith2020QRP}, confirmation or model-verifying frames occur when students view the goal of the lab as collecting data to confirm or support a known or provided physical model. In contrast, students holding an inquiry-based frame, which is more closely aligned with the instructional goals of these labs, view the lab space as a place to explore physical concepts by engaging in authentic experimentation \cite{SmithPERC2020}. Drawing upon prior work that identified students aims of finishing labs quickly \cite{Smith2020QRP}, we identified a third frame, hoops (as in ``jumping through hoops"). In this frame, students view the lab as a series of tasks they have to get through in order to leave the room. As will be shown, confirmation and hoops frames are not mutually exclusive: students may view confirming a model as a hoop to jump through.

Here, we present an analysis of the three non-problematizing groups, E, F, and G. 

\subsection{Phase 3 Findings}




\subsubsection{The Quiet Group E}
\label{AnalysisE}

The Quiet Group E consists of two students: Katie and Paulo. As identified in our Phase 1 analysis, the Quiet Group E spent the majority of their time in silence. Of the time we could determine they were speaking, about 40\% is inaudible. The primary time we could hear dialogue from both students occurred with the TA present at the table. Here, we present an analysis of one of their discussions with the TA. This is our only audible conversation where they clearly discuss some of their results. 

From snippets of audible speech, we can tell the group has difficulty collecting data initially. They first attempt to use the beach ball, but switch to the basketball, which they find easier to use. During this time, they do not ask for help from the TA. Around 35 minutes into the small group work, they receive some assistance from a lab manager who visits the lab classroom. They then successfully collect basketball data about 40 minutes into the group work time (80 minutes into the lab session overall). Later, the TA comes to their table and realizes they have not yet compared any data to the models. He directs them to do so, returning to their table 17 minutes later, when the following conversation takes place. 

\begin{quote}
\textbf{TA:} Alrighty, what did you, what have you guys got?\\
\textbf{Paulo:} Um, uh, we wrote that the um falsifying the model.\\
\textbf{TA:} Okay.\\
\textbf{Paulo:} But um [holding paper], the like revised experiment um, do we need, can we just test the beach ball to see if the uh.. uh if there’s, because there’s not much air drag on the basketball since it’s, it’s like, I don’t know, dense?\\
\textbf{TA:} Well, the basketball and the beach ball are about the same size and shape \\
\textbf{Paulo:} Yeah, the beach ball I mean.\\
\end{quote}

The TA asks about their findings thus far. Mirroring language the TA had used in his instructions, Paulo refers to ``falsifying the model," though it is unclear in this context what he means. He narrows the conversation to what they \emph{wrote} in the lab notes, focusing on the final product of the lab. He then turns his attention to a sheet of paper, which is likely the lab manual (hard copies of the lab manual are on the tables). His reference to the revised experiment is likely coming from this manual. He asks if they can ``just test the beach ball." This use of the word ``just" suggests Paulo is looking for a simple next step. He justifies this choice by implying he expects the different density of the two balls to cause different interactions with air drag. This justification comes after a pause and more halting, slower speech, suggesting he is unsure of these ideas or may be having these ideas for the first time. This indicates he was more sure of \emph{what} experiment to conduct than \emph{why} that experiment is justified. It is unclear what he means by his last statement referencing the beach ball. Overall, Paulo's focus on what should go in the lab notes and a straightforward revised experiment is evidence of a Hoops frame: He is trying to fulfill the requirements of the lab. 

The TA picks up on this framing and directly asks the students why they are planning to test the beach ball. The discussion continues:

\begin{quote}
\textbf{TA:} So um, why are you revising your experiment? What data did you see that, that, um, justifies your need to revise your experiment?\\
\textbf{Katie:} I (inaudible) \\
\textbf{TA:} I’m sorry I can’t quite hear you.\\
\textbf{Katie:} So I know that there’s air drag.\\
\textbf{TA:} Okay.\\
\textbf{Katie:} Force, one of the force applied on the object is air drag.\\
\textbf{TA:} Do you know that?\\
\textbf{Katie:} Oh, I assumed. \\
\textbf{TA:} Okay.\\
\textbf{Katie:} But um on our graph [turns laptop to TA], the acceleration was quite constant, and it was constant to the magnitude of ten which is similar to gravity. So it’s not measuring the effect of air drag, so we want to use the ball that might be more resistible (sic) to air.\\
\textbf{TA:} Okay, so your, you say your data um rules out the air drag one and seems consistent with the gravity only model.\\
\textbf{Katie:} Yeah.\\
\textbf{TA:} So um the advice I want to give you is to make sure not to design a confirmation experiment. 
\end{quote}

As the topic of the conversation switches, so does the speaker, with Katie now responding to the TA. Katie focuses in on air drag, which she assumes exists. She suggests that their choice of the beach ball is to specifically address the fact that their basketball data did not show the drag force they expected. Therefore, it appears possible they wanted to collect data with the beach ball to confirm that the drag force matters in some cases. The TA appears to pick up on this possible confirmation framing. After this moment, he encourages them to think of their revised experiment as about finding limitations on the gravity model rather than confirming a drag model.

Given the lack of audible speech, we looked at their lab notes to see if we could further understand how they framed their experimental choices. They state the purpose of testing the beach ball is ``To see whether the effect of air drag is marginal for other objects as well." While they do provide a graph of data from one trial of the beach ball in their notes, they do not analyse the acceleration on the way up and down separately. Instead, they assert that the low acceleration evident from the graph supports the air drag model. This claim was directly refuted in a full group discussion that took place before they collected their beach ball data.

Taken together, this is evidence that Katie is in a confirmation frame for portions of the lab: She expects to collect data that confirms the drag model. Paulo likely shared this framing at times as the lab notes were intended to collaboratively written and we see both students frequently typing on their laptops. 

Paulo and Katie's (the only two students in this group) speech appears to reflect two different frames. Given the flow of conversation, it is unclear if this is a shift in frame, if they had been framing the revised experiment differently throughout, or if these frames are overlapping (seeking to confirm a model is a potential way to complete a task quickly). As mentioned in our Phase 1 analysis, Katie and Paulo sit on opposite ends of the table, often in silence. This lack of interaction may either reinforce or be reflective of differences in framing. Due to the lack of other audible discussion, we are unable to infer their framing during other discussion of data.


\subsubsection{The Uncertain Group F}
\label{AnalysisF}

The Uncertain Group F consists of three students: Michelle, Ariel, and Kyle. From the Phase 1 analysis, this group looks more like the four tables that engage in problematizing around the intended problem than they do the non-problematizing Group E: they are consistently engaged in on-topic conversation and they obtain and discuss the beach ball early in the lab session. What off-topic conversation they do have emerges from discussing the lab: after trying to throw the balls so that the motion detectors collect data, Michelle and Ariel pivot to a conversation about how they observed tennis balls moving and they discover a shared hobby. 
Yet, the Uncertain Group F never meaningfully discuss the possible causes of the discrepancy between their data and the beach ball data. In contrast to groups that successfully problematize, this group remains in a stable confirmation frame throughout the lab session and does not take up the epistemic agency for which the lab makes space.

At the beginning of the lab, the group decides to measure the acceleration of a beach ball. Their first trial results in an acceleration of 6.94 $m/s^2$ and they fixate on the correlation coefficient of the applied linear fit. Without discussing a reason why, the group discards this trial and takes a few more measurements. They then ask for help from the instructor, who tells them they have collected reasonable data from which they should be able to draw conclusions. They look at the most recent trial and compare their results to the given models:

\begin{quote}
\textbf{Michelle:} Our velocity equation is negative $6.792 t + 9.558$ with an R squared value of negative $0.999$ so basically 1 which is good. So if we want to do $a$, we’d do negative 6.792. \\
(4 second pause)\\
\textbf{Michelle:} So I mean, it’s because it shouldn’t be 8 because there’s drag, right?...But it’s like close enough to 7 that it should be fine, right?\\
\textbf{Ariel:} [laughs]\\
\textbf{Kyle:} Yeah.\\
\textbf{Michelle:} Because if there were no drag there’d be 9.81, but it's within that range where I think it makes sense?\\
\textbf{Kyle:} Okay.\\
\textbf{Michelle:} If you look at the um, if you look at the free body diagrams you have like drag (inaudible) that means it's the same direction and magnitude for your acceleration vector, but it's going less than. So I think we're good.
\end{quote}

As Michelle verbalizes, they find a constant acceleration of $6.792 m/s^2$ for the beach ball's flight (as one fit including both up and down), which violates the predictions of both models. She states that it is a reasonable value to get based on the model that includes drag (Model 2). She claims that while the gravity only model (Model 1) predicts a magnitude of acceleration equal to $9.81 m/s^2$, the second model should predict an acceleration value lower than $8 m/s^2$ due to drag. Because their value is ``close enough to seven,'' she concludes that their measured acceleration supports the second model. Michelle evaluates the $r^2$ value of $-0.999$ as ``good," because the value indicates a close linear fit (they do not distinguish between the motion upwards and downwards).  Michelle's evaluation of the data as ``good" and ``close enough" indicates she is seeking to obtain data that fit certain expectations, evidence she holds a confirmation frame. 

The instructor then comes to talk to the group and advises they ``look in more detail about what these models predict and if that happens throughout the trajectory of this falling ball," alluding to the difference between the acceleration up versus down.  Despite explicit instruction to think more critically about their results, the group subsequently defers to discussing what to write about in their lab report, after which they ask the instructor for help. They drop their discussion of data and ignore  the instructor's bid and instead focus on what task they should complete next. These actions may indicate a shift to a hoops framing--they view fulfilling the requirements for the graded notes as an important task.  When the TA returns, Michelle asks the TA about how their data fits with the drag model (Model 2).



\begin{quote}
\textbf{Michelle:} I thought it was close enough to like assume that there was some drag because it would be less than 9.81 if there is some drag.\\
\textbf{TA}: What you've done is, okay, this is important, you gotta get in the falsification mindset. You're still looking for what it does say, what you need to consider is what does it rule out. So you didn't say this confirms drag, what you really did was ruled out the gravity only model. And there are features of it that are consistent with drag, but when you are comparing model to data, really what you're only looking is for (sic) what it eliminates as a possibility.
\end{quote}

Again, the group frames their results as ``close enough'' to their expectations that they can make claims about each model's validity. In response, the instructor makes an explicit attempt to confront their confirmatory frame and encourages them to rethink their conclusions. 
Later on, they talk to the instructor again and discuss the shape of their acceleration graphs, which only leads them more toward confirming Model 1:

\begin{quote}
\textbf{TA:} What shape is this?\\
\textbf{Michelle:}  It's a linear line.\\
\textbf{TA:} It's linear. So what does that mean, so if the velocity is linear what's the acceleration?\\
\textbf{Ariel:} Oh it's constant acceleration. Oh but we expect it to not be constant.\\
\textbf{TA:}  Well, yeah you expect it to not be constant, but the, you mean Model 2--\\
\textbf{Ariel:}--Yeah--\\
\textbf{TA:} (continuing)--predicts it wouldn't be constant. Yeah which is also kinda what you expect.\\
\textbf{Ariel}: Yeah. (laughs)\\
\textbf{TA:} So now, I recommend you draw a free body diagram that explains the data you see. You've got a free body diagram for each of these models. Now draw a free body diagram for what is actually happening. Okay?\\
(TA motions to leave table) \\
\textbf{Michelle:} Wait, so even if there was drag, there would still be constant acceleration?\\
\textbf{TA:} Here? That's not what that predicts.\\
\textbf{Michelle:} Because this is sounding more like this, but then like the problem is that it's not closer to 9.8 at this point because it's constant. \\
\textbf{TA:} So that's what you need to figure out. That's the challenge that's placed on you today. That's why I said you'll find something weird. I warned you. I told you it would happen. That's why we're doing this and talking about confirmation bias at the same time. Because that's--it's weird!\\
\textbf{Ariel:} Alright\\
\textbf{TA:} So I can't tell you what to do about it or else it ruins the whole thing but you are correct in noticing--you found the right weird thing. You have succeeded in finding the weird thing which not everyone has done yet.\\
\textbf{Ariel:} Yay.\\
\textbf{TA:} So keep going. Draw a free body diagram and try to explain it.\\
\textbf{Ariel:} Okay.\\
(TA leaves table)\\
\end{quote}

The TA's questioning results in Ariel realizing that their data is inconsistent with their expectations for how the ball should behave if drag has a significant impact on the motion. Ariel's statement that they ``expect" non-constant acceleration is evidence of a confirmation frame. The TA builds on her words and highlights what Model 2 (the model with drag) \emph{predicts} and offers explicit direction to draw a new free body diagram, moves which appear to be an attempt to push the students away from their confirmation frame. He also highlights that the students have found ``the weird thing" and that this is a success. He ends his interaction with the group reiterating the instruction to draw a new free body diagram.

Just seconds after the TA leaves the table, Ariel refocuses on what the TA said about Models 1 and 2. 

\begin{quote}
\textbf{Ariel:} So basically, wait--did he say with model--he say (sic) you'd expect non-constant acceleration?\\
\textbf{Michelle:} Yeah so with Model 1 is constant.\\
\textbf{Ariel:} So Model 1 is constant?\\
\textbf{Michelle:} Model 1 is constant Model 2 is non constant.\\
\textbf{Ariel:} And we found constant but not the right value.\\
\textbf{Michelle:} Yeah.\\
\textbf{Ariel:} Alright, I'll write that down.
\end{quote}

The students do not follow the TA's instructions to consider a new model by drawing a new free body diagram. 
Instead, they return to their discussion of the given models.  They appear to infer that rejection of Model 2 implies that Model 1 is correct. Though Model 1 does predict a constant acceleration, the magnitude of the group’s measured acceleration is significantly below the predicted value. Therefore, they conclude that their data support Model 1, but ``not the right value." 

Later, a member of another lab group asks the group what they have seen so far, to which Ariel responds, ``we got that it’s a constant acceleration but that it's less than 9.8 so that it rules out both models.'' In passing, they draw the correct conclusion based on their data, but once the other member leaves, the group immediately defaults to thinking about how to alter their data in favor of Model 1: ``what do we do to like make it closer to 9.8?" This is an implicit bid to redo the analysis or data collection for the purpose of obtaining the value that would confirm the model and even more evidence of their stability in a confirmation frame. Rather than engage with the source inconsistency between their data and models, Michelle's bid is to engineer data to better fit her expectations.

After the instructor announces there are 35 minutes in the lab session left, the group decides that they need to perform a ``revised experiment" as instructed by the lab manual. They enter a hoops frame in which they are 1) stymied by the fact that their initial data does not confirm either model and 2) unsure of how to use their findings to generate a follow-up exploration. They default to repeating the same measurements but with a basketball instead of a beach ball. When considering what to report in their lab notes about justification for their new experiment, the group debates possible explanations for their choices for the sake of the assignment. In the discussion that follows, Ariel is sitting at her laptop and pauses from writing lab notes to make her statement.

\begin{quote}
\textbf{Ariel:} So we’re thinking either rotational or something to do with the air pressure inside the ball.\\
\textbf{Michelle:} Yeah, I like--\\
\textbf{Ariel:} But do you have an idea, do you prefer one of them? Or do we have more justification for one of them? Like, I don’t know?
\end{quote}

Although they are able to come up with a few factors that could be causing the disagreement between their data and the models, such as a ``rotational" force or ``air pressure," the group does so only after taking the new data. For the first time during the session, the group discusses physical forces acting on the ball and offers explanations for their earlier discrepancies. Neither idea is discussed with respect to their data. Ariel does make a weak bid for providing some link between their data and these forces by asking if there is ``more justification," however her immediately preceding question implies that simply preferring one would be a satisfactory reason for choosing an explanation. This is further evidence of a hoops frame, where Ariel is deciding what to write in the notes rather than engaging with the underlying uncertainty.

They ultimately propose that ``whatever the [unknown] force is, right, it has to be changing at the same rate as...drag," so air pressure changing with altitude makes sense to investigate. The group drops a basketball close to the ground and reacts positively to their results:

\begin{quote}
\textbf{Michelle:} Oh! It's -9.607.\\
\textbf{Ariel:} Yeah!\\
\textbf{Michelle:} For this line.\\
\textbf{Ariel:} Yes!\\
\textbf{Michelle:} Then let's do this line, too. (10 second pause). No, get off. This is -9.209...this is -8.517 (7 second pause) this is 10.2, 10.1...-10.29, so that's a little high.\\
\textbf{Ariel:} We said -6.792 is what we found for acceleration, right?\\
\textbf{Michelle}: Yeah. (10 second pause) This is -8.936, okay so we are like closer to 9.81\\
\textbf{Ariel:} Okay so maybe...\\
\textbf{Michelle:} So, what does that mean?\\
\textbf{Ariel:} That, so are we saying it's air pressure, that air pressure changes, so...\\
\textbf{Michelle:} We could say that, let's ask him [the TA].\\
\textbf{Ariel:} Okay.\\
\textbf{Michelle:} [TA] we need your help!
\end{quote}

The group expresses excitement when their measured acceleration is close in magnitude to the gravitational acceleration, as predicted by Model 1. They judge their data by how close it is to what they expect, claiming 10.29 $m/s^2$ is ``a little high," evidence they are still in a confirmatory frame. When they try to make claims about how air pressure impacts the ball's motion, they again resist the agency they have in making interpretations and conclusions by deferring to the instructor. At the end of the lab period, the group decides to take more data and compare accelerations for the basketball dropped at two different heights, to see if air pressure at different altitudes is contributing a force not accounted for in the two models. Their statistical analysis unveils that the two accelerations are indistinguishable, after which they draw their final conclusions:

\begin{quote}
\textbf{Michelle:} Because the \emph{t} value \footnote{The lab materials use ``t prime" as a short hand for for the output of a T-test to evaluate if two data sets are distinguishable.} is so small, we can do a model that doesn't account for the pressure force.\\
\textbf{Ariel:} Since the \emph{t} value is so small... \\
\textbf{Michelle:} Yeah. You can do a model that doesn't account for the pressure force, cuz like remember how the pendulum the t-value was like above 3.\\
\textbf{Ariel:} Mhm.\\
\t'xtbf{Michelle:} So like you had to account for angular displacement, because this one, because it’s so close to 1 we can pretend like its negligible--\\
\textbf{Ariel:} --Okay--\\
\textbf{Michelle:} (continuing)--if we wanted to.\\
\textbf{Ariel:} So it’s negligible, so it’s basically Model 1?\\
\textbf{Michelle:} Yeah.\\
\textbf{Ariel:} (typing) Okay, so we can say like thus, our results--\\
\textbf{Michelle:}--Our results show that Model 1 is probably the best...estimation.
\end{quote}

By the end of the lab, Michelle and Ariel hint at the possibility of a third force not accounted for in either model that may be influencing their results. They even compare this train of thought to the previous lab unit, in which they explicitly learned that physical models sometimes make assumptions or simplifications at the surface level. Yet, their final statements reflect a return to their earlier confirmation frame, in which they decide that Model 1 offers a good physical description of the phenomenon of interest irrespective of the data they collected. This group’s inability to break out of a confirmatory stance toward the lab task, despite repeated attempts by the instructor to push them out of it, prevents them from engaging with the inconsistency and thus problematizing. The TA was aware that this group struggled: as they packed up, he noted that their struggle was ``productive" as they were grappling with the data and told them that he made an active choice to not give them more guidance. The students made attempts to process what the TA had told them do do and their active, cooperative engagement in the lab activity strongly suggests they were trying to do what they are supposed to do. Notably, Kyle, the male student in the group, barely speaks during these discussions despite always being present at the table.

\subsubsection{The Social Group G}
\label{AnalysisG}

The Social Group G consists of three students: Rachel, Brett, and Mike. In contrast to the Uncertain Group F, Group G obtained usable data from the beach ball latest in the lab session--their first attempt at collecting data with the beach ball was unsuccessful--and they did not immediately discuss the data in detail. A half hour later, while writing the lab notes they finally discuss the data. They appear to have a hoops frame: their shared understanding of what is taking place is that they are jumping through a set of hoops to complete the requirements. 

When they initially collected data from the beach ball, only Rachel and Mike are at the table. When Brett returns to the table, their topic of conversation shifts. Therefore, they do not discuss the beach ball data until there are only 12 minutes left in the lab session and they only do so in comparing their results between the two balls.
During this time, Rachel is seated and typing on her laptop, while Mike and Brett switch between seated and standing positions, occasionally looking at the desktop computer that was used for data collection.  

\begin{quote}
\textbf{Rachel:} So it's [the acceleration of the basketball is] indistinguishable going up and down. \\
\textbf{Mike:} (mumbling) It looked like that\\
\textbf{Brett:} So that, so that one is very similar to Model 1. No air drag.\\
\textbf{Mike:} What does he [the TA] mean by show when it fails? \\
\textbf{Brett:} When the answer is that Model 1 holds with uh the basketball however it fails with the beach ball...um... so there must be some. So then-- \\
\textbf{Mike:} --We can't say it's air drag\\
\textbf{Brett:} No no no exactly. We've come to the conclusion that as a result of--\\
\textbf{Rachel:}--so what--\\
\textbf{Brett:} (continuing)--something to do with the ball, differences among the ball are causing differences among results. \\
\textbf{Rachel:} (typing) Well so we used uh the basketball this time instead of a beach ball to investigate if perhaps something other than air drag is responsible for the difference. Um. (shrugs)\\
\end{quote}

While they acknowledge their data is inconsistent with both Model 1 (the gravity only model) and ``air drag," they discuss their findings in terms of ``the answer" and stating that ``we've come to the conclusion," indicating that they are not framing this conversation as an active discussion of the meaning of their results. Because Mike's statement of ``We can't say it's air drag" comes as Rachel is writing in the notes, he seems to focus on fulfilling the notes requirements rather than whether or not air drag explains their data. While they acknowledge the differences in acceleration are due to differences among the balls, they do not engage in a discussion about what those differences might be. Taken together,  their conversation reflects a frame of jumping through hoops to complete the lab in the allotted time. This framing makes sense because they are running low on time to complete their investigation.

After a minute of silence, their discussion continues.  At times, Rachel speaks softly while typing on her laptop and is not audible. 
\begin{quote}
\textbf{Rachel:} Wait we never answered the question ``quantify the level of agreement between the models and our measurements"?\\
(6 second silence)\\
\textbf{Mike:} I mean isn't that what the statistical analysis that we did showed? Because our models, like, the models suggest what it should be and the statistical stuff will tell us how accurate ours are...\\
\textbf{Rachel:} But the statistical stuff shows how close they are to each other. \\
\textbf{Mike:} Yeah but that's like kind of a measure of accuracy. \\
\textbf{Rachel:}  (inaudible)\\
\textbf{Mike:} I don't know. \\
\textbf{Rachel:} Um...We calculate the expected value and compare that to the value in our data set. \\
\textbf{Mike:} Compare how?\\
\textbf{Rachel:}  (inaudible) \\
\textbf{Mike:} Sure. I mean last time he talked about that and he only took off a quarter of a point. So I feel like he doesn't really care about that.\\
\textbf{Rachel:} I feel like there's a lot of lab notes and five pages each. 
\end{quote}

Rachel notes that she thinks they missed one question asked in the lab manual. Mike argues that they fulfilled that requirement already. Though it's hard to hear the intervening conversation, Mike highlights that not answering that question did not result in a significant deduction on the previous set of lab notes. Rachel appears to concur, noting that the TA has many lab notes to read and grade. Their primary attention is on how to write notes efficiently to satisfy the TA. Thus, we describe them as continuing in their hoops framing. 

Four minutes later, with less than ten minutes left in the lab session, they talk to the TA about what they found. They appear to initially be in a confirmation framing as they present their findings to the TA:

\begin{quote}
\textbf{Rachel:} Okay so while we found with the basketball that it was 9.5 on the way up and like 9.52 on the way down.\\
\textbf{TA:} Mmhmm \\
\textbf{Rachel:} and so the $t$ prime between those is like 0.37, so it's probably similar. \\
\textbf{TA:} Okay so for the basketball, it's the same up and down. What does that mean? \\
\textbf{Brett \& Rachel:} Um...\\
\textbf{Rachel:} That it will be Model 1. \\
\textbf{Brett:} Yeah, that Model 1 is very applicable when they're... \\
\textbf{TA:} So, close. You found that model two \emph{isn't} applicable. \\
\textbf{Mike \& Brett together:} Yeah.\\
\textbf{Brett:} Okay. \\
\textbf{TA:} That's different.\\
\textbf{Brett:} Yeah. \\
\textbf{Rachel:} Oh.\\
\textbf{TA:} Or at the very least, you could quantify the extent to which model one, you could say for the basketball, Model 1 predicted the result for this many decimals, Model 2 predicted it to that many and, you know, Model 1 is better.  \\
\textbf{Brett:} So our final answer is that model, the basketball demonstrated that Model 2 didn't hold then yet the beach ball showed that Model 1 does not hold.  \\
\textbf{TA:} Okay. So the fact that there's a difference between them is an interesting result that you can in fact investigate. And so you'll do that more next time, you'll say, ``What is going on, why these are different, which forces explain that?" \\
\textbf{Brett:} Got it. 
\end{quote}

Their initial comment is that their basketball data are consistent with Model 1 because the acceleration is indistinguishable on the way up and down. They focus on being able to apply Model 1, rather than reject Model 2. The TA corrects them and notes that they have not done enough analysis to support how well their data agrees with Model 1. Brett rephrases to again contrast the basketball and beach ball findings, describing their ``final answer" which rejects both models using different data sets. As the TA elaborates that they will investigate this more in the next lab session, Brett responds with a short ``Got it" while Rachel continues to type on her computer. While they were inclined to think of their findings in terms of confirming a model, with so little time left in the lab, Brett is willing conclude what he believes the TA wants him to conclude. This hoops framing makes sense at this point in the lab: shortly after speaking to this group, the TA tells groups to wrap up their work quickly so that the
room will be available for the next lab session. 

Overall, throughout Group G's discussion of the meaning of their beach ball data, we see them in a hoops framing, with some possible moments of confirmation framing. 


\section{Discussion}

Through our three phases of analysis, we documented how large scale behaviors, interactions with the teaching assistant, and framing differ across the groups. In Phase 1, we identified that though the majority of all groups' speech was on topic, the audible groups that did not problematize tended to have more off-topic conversation than those that did problematize. We further saw that Group E was an outlier with a large fraction of time spent in silence. In Phase 2, we saw the number of times the TA interacted with different groups was relatively consistent with no clear pattern between groups that did and did not problematize. A primary difference in the TA's interactions was that the TA spent far more time discussing interpretation of results with Group F, who did not problematize. Our framing analysis in Phase 3 provides some explanations for the differences we see in the large scale behaviors and TA interactions: the non-problematizing appear to frame the lab as about confirming models and jumping through hoops. 

We now discuss each of the three groups that did not problematize around the intended problem (Groups E, F, and G) individually, looking across the different phases of analysis.

\subsection{Summary of findings by group}

\subsubsection{The Quiet Group E}

Given the poor audio quality when the TA was not present at the table, the analysis we were able to perform with Group E was less complete: we could not reliably identify when they were on or off task and could not identify when they discussed their beach ball data that should have given rise to the intended problem. Our Phase 1 analysis allowed us to see a broad lack of engagement with the laboratory tasks: they do not ask for help even when they appear to struggle with data collection, they spend far more time in silence than the other tables, their lab notes are substantially shorter than other groups, and they frequently sit separately from each other with their gaze on their individual devices. In what we can hear on the audio and read in their lab notes, we do not see any evidence they confronted the issue of the unexpectedly low acceleration of the beach ball. The TA spent somewhat less time with Group E than most other groups and spent very little time discussing the interpretation of results with them, and we believe the TA largely did not notice their lack of engagement. When the TA does discuss their basketball data with them, Paulo's discussion of their plans for their next step reflects a Hoops frame, focusing on what is required in the lab notes and a simple next experiment. Katie's justification for their revised experiment reveals a possible confirmation frame. This suggests that along with their overall lack of engagement, Paulo and Katie may be working in parallel, rather than together, for much of the lab.

\subsubsection{The Uncertain Group F}

When beginning this work, the Uncertain Group F puzzled us the most: they collect clear data that shows the intended problem and begin discussing it earlier than the other groups that did not problematize. They are consistently talking to each other and on task. The TA spends more time with them than any other table and spends over half of that time interpreting results with them. This is likely related to their framing: the TA makes efforts to push back on their confirmation framing, particularly while interpreting results. While they follow his suggestion to not take additional data, they do not take up the inquiry framing the TA encourages. Overall, this group appears to not see themselves as epistemic agents \cite{stroupe2014examining}: they understand lab as a place to confirm known results and not to construct their own knowledge. While they earnestly attempt to complete what is required by carefully reading the lab manual repeatedly, their stability in their confirmation framing for much of the lab prevents them from perceiving opportunities to construct new models or understandings. The TA does not meet that expectation and is aware that he is not meeting their expectation of telling them what to do and what to conclude. In response to this framing and agency mismatch, the students sometimes switch to framing the lab as simply a set of hoops to jump through. For this group, the hoops appear to be defined by the lab manual, as they frequently attend to the lab manual's text while ignoring instructions from the TA.

\subsubsection{The Social Group G}
\label{SocialGroupAnalysis}

The Social Group G collects data relevant to the designed problem later than other groups and they do not discuss that data until very late in the lab period. Several things may have contributed to this late discussion: Because Brett frequently moved around the room talking to other groups, the topic of conversation sometimes shifted abruptly, as happened when they first collected their data. Their off-topic conversation may also have slowed down their progress, contributing to them focusing on finishing their report late in the lab. In that context, it is logical for them to take on a hoops framing: they want to ensure they submit complete lab notes by the deadline. Unlike Group F, who closely attend to the lab manual, Group G's focus is on how the TA will grade the notes. It is not clear whether or not the nature of the off-topic conversation--which was far removed from the lab context--impacted their engagement. However, the statement by Rachel that she felt like she was in a frat suggests that she was at least somewhat uncomfortable in her group.

\subsection{Understanding what contributed to students' lack of problematizing}

Studies that have examined students' engagement in reformed labs have shown some mixed results: while these labs have great potential to engage students in a broad range of scientific practices \cite{Brewe2008, meyer2017student, Zwickl2015, Dounas-Frazer2016, HolmesPNAS, Bartlett2019}, their engagement is not always productive 
\cite{SmithStienHolmes2018, SteinPERC2018, Smith2020QRP}. 
In many ways, the lab session studied here was quite successful: of seven student groups, four arrived at the intended problem and engaged with it productively \cite{ProblematizingPERC}. Of the three groups that did not, we do not see evidence that they took on an inquiry framing that is necessary for problematizing when discussing their beach ball data: the students did not approach their data as containing an interesting question to resolve. In particular, Group F does not act with epistemic agency, despite the TA's repeated moves to encourage them to think critically about their data and models and draw their own conclusions. While it seems possible that Group G may have engaged differently had they collected their beach ball data earlier or that Group E might have engaged more deeply with the lab with additional intervention, Group F's stability in a confirmation framing seemed to fully block their ability to engage in productive problematizing despite explicit intervention from the TA.

Understanding that managing students' uncertainty in the classroom requires raising, maintaining, and reducing uncertainty  \cite{gouvea2020argumentation, chen2019managing}, we can see that the TA raised and maintained uncertainty with both Group F and Group G. However, it is possible that managing a group reluctant to take on a role as epistemic agents requires not only attending to their framing \cite{hutchison2010attending}, but also a careful balance of reducing uncertainty for groups who struggle with making progress when their data is inconsistent with or pushes back on their original understandings \cite{manz2015resistance}. It is conceivable that had the TA spent \emph{less} time trying to push the students in Group F out of a confirmatory framing and offered more direct instructions for how to proceed, the students might have been able to have more productive discussions about the underlying problem.

Our analysis of the groups' on- and off-topic discussion also suggests that the group social dynamics may contribute to students productively taking on productive problematizing. In particular, the Social Group G in particular spent much more time off-topic, dispersed off-topic discussion throughout the lab, and engaged in off-topic discussion that may have been uncomfortable for some members of the group (i.e., alcohol and spring break parties). With this group, we particularly note the possibility of gender dynamics impacting the group productivity. Research has extensively demonstrated the differential and often inequitable ways in which male and female students participate in hands-on lab activities~\cite{quinn2020group, QuinnPERC2018, doucette2020hermione, day2016, danielsson2014physics}. The dynamics between the female student and one of the male students reflect the archetypes of \textit{Hermione and the Slacker} from Doucette et al.~\cite{doucette2020hermione}. This male student is also absent for much of the lab, regularly circling around the lab room visiting other groups. His absence may have limited the progress the other two students could make and, when he was present, much of his time was off-topic. We also see that the two female students in the Uncertain Group F dominate the conversation, both on- and off-topic while the male student barely speaks. The equity or inequity of group discussion may impact students' engagement in productive practices, such as problematizing.

Finally, it is unclear from our analysis what caused the Quiet Group E's overwhelming lack of engagement (as evidenced by their significant time in silence). Without access to more of their conversation, we cannot accurately infer how their framing or the role of social dynamics impacted their engagement. This presents an important limitation of the observational nature of our work: there is much about students' framing and understandings that is unobserved when analyzing video and audio transcripts. Future work should evaluate the ways in which other forms of evidence, such as students' written work, can provide glimpses of students' frames or dynamics. Written work has been used effectively to evaluate students' engagement in scientific practices~\cite[e.g.,][]{SteinPERC2018, Smith2020, HolmesPNAS, etkina2000weekly, Etkina2006,Etkina2008, Etkina2010, stanley2017using, gouvea2020argumentation, elliott2016composing}, suggesting it may also be informative for inferring framing.

\subsection{Implications for instruction}

Our previous work~\cite{ProblematizingPERC} is consistent with that of other researchers who show that designing uncertainties into science classrooms can lead to productive engagement in science practices \cite{manz2018supporting,manz2018designing, ko2019opening, chen2020dialogic}. The current study provides further evidence that such work is challenging for instructors. The TA for this lab session had physics teaching experience in a variety of contexts and had also taught this set of labs before. He agreed with the goals of the lab and worked to support students' framing the lab as a setting to uncover and investigate questions. It is quite possible that had we conducted our study in a session taught by a novice TA or a TA with traditional views of lab instruction, our results would be quite different. It is also possible that students' framing and engagement in problematizing shifts over time. Here, we looked only at one lab session, the fourth of ten total. Students' failure to problematize in the first half of the semester may be a type of ``productive failure" \cite{kapur2008productive} that sets them up to be more successful in later labs. Understanding students' framing and problematizing over longer time scales will aid in determining how to influence their framing.

Our work also supports that teachers may need specific training to support students' productive engagement with uncertainty \cite{ko2019opening, chen2020dialogic}. In particular, TAs may benefit from specific training on balancing how they both maintain and \emph{reduce} uncertainty for students \cite{chen2019managing,gouvea2020argumentation}. Because students' problematizing may be similar across age ranges and contexts \cite{phillips2017problematizing}, we expect that work done in K-12 science education research in supporting teachers in engaging with student uncertainties applies to college settings. In particular, attending to the issue of supporting students' epistemic agency may provide a framework to help TAs open up the space in their laboratory classrooms \cite{chen2019managing,ko2019opening}. This work is not without challenge: As Ko and Krist note, ``Positioning students as epistemic agents requires an intentional redistribution of power" \citep[p. 981]{ko2019opening}, and students and instructors in all contexts may resist that distribution of power. If a TA is reluctant to shift power from themselves to students, it is unlikely they will encourage students to take up epistemic agency as firmly as the TA in this lab session.  This challenging aspect of TA professional development deserves further attention in research. 

We also see more straightforward challenges the TA faced in this lab: the TA did not appear to realize how broad Group E's lack of engagement was and he stated at the end of the lab that he did not realize that Group G did not delve into the beach ball data earlier. Laboratory TAs, and instructors in general, often face the challenging teaching task of managing many groups of students at once. It is unsurprising that any instructor, even a very experienced one, may miss important aspects of some groups' engagement in the lab tasks. 

Furthermore, we can see students weighing cues about how to participate in the lab: the Uncertain Group F attends closely to the lab manual, and the Social Group G openly states that the TA may not take off many points for failing to include adequate error analysis. In these ways, we can see these students attempting to be successful: they are trying to fulfill the stated and implicit requirements of the laboratory task. Careful alignment of materials, TA training, and assessment may aid in helping students' frame labs in the way that curriculum designers intend. Indeed, students in this study seek to justify their revised method due to a prompt in the lab manual. As such, we can interpret their hoops framing as leading them to engage in an intended practice. Critical evaluation of the materials may help identify when hoops framing may be productive or unproductive.




\section{Conclusion}

By comparing seven lab groups in a single lab session, we have identified key features that appear to separate groups that did problematize around an intended problem from those that did not. Students' engagement, framing, and epistemic agency are critical: if the groups are not meaningfully engaged in a task, like the Quiet Group E, or if they take on a confirmation or hoops frame while discussing their data, like all non-problematzing groups, they may not problematize. For the audible non-problematizing groups, we can see that their actions make sense: the Uncertain Group F spends time trying to understand the lab manual and confirm the given models, while the Social Group G has to finish their work quickly. 

As four out of the seven groups problematized around the intended inconsistency, our present and prior work \cite{ProblematizingPERC} is consistent with science education research at the K-12 level that notes curricula can be designed and enacted to enable students to act with epistemic agency \cite{ko2019opening, manz2018supporting}. The current project also highlights the difficulties TAs may face in supporting students' problematizing in lab settings, but uncovers potential suggestions to suppress them: TAs of these kinds of labs may benefit from additional support for viewing students as epistemic agents. 

While we see students' framing as important in their engagement in the lab, future work must be conducted to address how and why students' framing may shift in the lab and the instructor's role in those shifts. Given that there is some evidence that previous experience in courses and labs that promote epistemic agency impacts students' actions in future courses \cite{SmithPERC2020,AnnaDiss}, understanding how to support students' stable framing of inquiry and epistemic agency is a key target for future research.


\bibliography{Problematizing.bib}

\section{Supplementary Material}
\label{Supplement}

\subsection{Laboratory Unit}
In this lab, students first conduct small and large group discussions on confirmation bias and brainstorm ways to mitigate it. They are then asked to predict the acceleration of an object using two models: one with only gravity and one with gravity and drag, as shown in Fig. \ref{fig:Models}. They are given beach balls, basketballs, and coffee filters to test, along with ceiling-mounted Vernier motion detectors. A full description of this lab is \hyperlink{https://www.physport.org/curricula/ThinkingCritically/}{available for instructor download at PhysPort}. (This is Mechanics Lab week 4 in those materials.)

\begin{figure*}[h]
    \centering
    \begin{subfigure}{3.5in}
        \centering
        \includegraphics[width=3.5in]{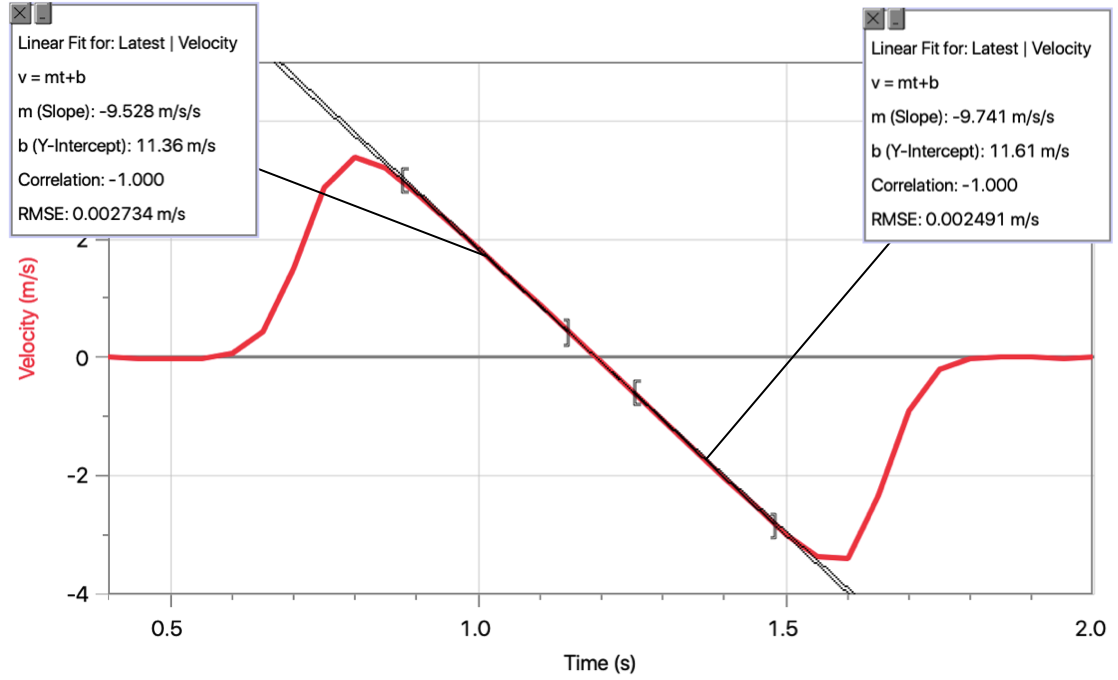}
        \caption{}
    \end{subfigure}
    \begin{subfigure}{3.5in}
        \centering
        \includegraphics[width=3.5in]{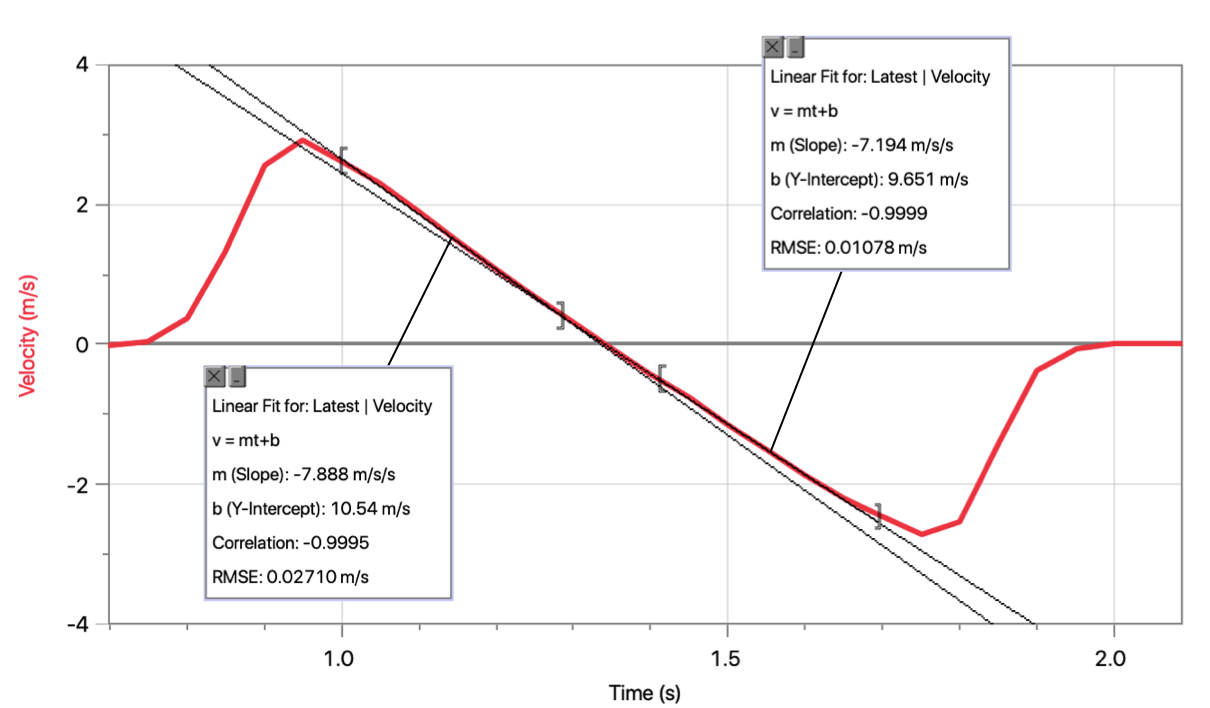}
        \caption{}
    \end{subfigure}
    \caption{Sample velocity data from (a) a basketball and (b) a beach ball, thrown upwards. Students use the linear fit function in LoggerPro to measure the acceleration upwards (positive velocity on graph) and downwards (negative velocity on graph). }
    \label{fig:Graphs}
\end{figure*}

The students have the choice of what object to test, however, in this session, all students focused on the two types of balls. Sample data similar to what is normally collected by students is shown in Fig. \ref{fig:Graphs}. Students are asked to compare their data to the two given models. The manual offers students the following guidance: 

\begin{quote}
$\bullet$ If you cannot distinguish the two models, work with your group to design an improved measurement method to better distinguish them. \\
$\bullet$ If you can distinguish the two models, repeat your procedure with another object and/or design a new experiment to further test the favored model. The better model may still be limited and there may be an even better model! 
\end{quote}

As with this sample data, students typically find that the acceleration up and down do not differ substantially for each ball. However, it is also typical to find acceleration of the beach ball that is significantly lower than the values predicted by either model due to the influence of the buoyant force as shown in Fig. \ref{fig:Models}. The lab manual states that students may need to consider another model that better matches the data. 

By the end of the lab session, all seven groups have collected and performed some analysis on data that shows the beach ball's acceleration is less than \emph{g} and the basketball's acceleration is near \emph{g}, with most also finding that the acceleration is approximately constant through the trajectory.

\subsection{Supplementary Coding Methodology}
\label{SupplementCoding}

\subsubsection{Phase 1}

\emph{Off-topic example coding}
\vspace{1pc}

The coding for Phase 1 was largely straightforward with little room for interpretation beyond on- and off-topic speech. All transcript in our framing analysis was coded as on-topic. An example of speech that was coded as off-topic is below.

Immediately prior to this, a student was looking for a calculator, asking his groupmates where it is. Those conversations were coded as on-topic, as the student was looking for the calculator to complete a relevant task. Once he finds it, a few seconds were coded as silence before his groupmate asks the following question: 
\begin{quote}
\textbf{Student 1:} Hey, where'd you get that?\\
\textbf{Student 2:} [University Name] Store.\\
\textbf{Student 1:} How much was it?\\
\textbf{Student 2:} Twenty bucks. Yeah because we can't have it for this class--we can't have a graphing calculator.
\end{quote}
There is another pause (coded as silence), and then another group member asks ``so what does that say?" while pointing to the computer. This speech, and the conversation about their data that follows, was coded as on-topic.  
\vspace{1pc}

\emph{Inter-rater reliability details}

\vspace{1pc}

Because our coding was continuous and codes were applied for various durations, we provide multiple ways of understanding the high interrater agreement. When sampling the data \emph{by time}, we find exceptionally high interrater agreement. We calculated Cohen's $\kappa$ using the built in tool in BORIS, which samples at fixed times. Using sampling times between 0.5 and 5 seconds, we obtain $0.865>\kappa>0.849$. This demonstrates that at any given moment, the two coders are highly likely to be in agreement.

A summary of code frequencies and total duration is in Table \ref{tab:Interrater}. The most common source of disagreement is whether or not the group was inaudible or silent. This is the primary source of the larger total time Coder 1 coded as inaudible. ``Non-overlapping" for each column in the table below mean instances where one coder applied a code when the other coder did not. For example, Coder 1 identified 8 instances of audible speech while Coder 2 identified 7. This is because Coder 2 marked 5 seconds as inaudible in the middle of a block of otherwise audible speech. Therefore, this moment contributes one instance of non-overlapping coding: one inaudible code for Coder 1. 

\begin{table}[h]
    \centering
    \begin{tabular}{|l|c|c|}
    \hline
         & Coder 1 & Coder 2  \\
         \hline
      Off-topic, number of instances   & 1 & 1\\
      Off-topic, total duration & 14.5s & 10.7s\\
      Off-topic, non-overlapping instances & 0 & 0\\
      \hline
      On-topic, number of instances & 6 & 6 \\
      On-topic, total duration & 421.6s & 414.9s \\
      On-topic, non-overlapping instances & 0 & 0 \\
      \hline
      Audible, number of instances & 8 & 7 \\
      Audible, total duration & 425.5s & 425.9s\\
      Audible, non-overlapping instances & 0 & 0 \\
      \hline
      Inaudible, number of instances & 6 & 2 \\
      Inaudible, total duration & 37.8s & 14.2s\\
      Inaudible, non-overlapping instances & 4 & 0 \\
      \hline
      Silence, number of instances & 8 & 6 \\
      Silence, total duration & 26.3s & 28.3s\\
      Silence, non-overlapping instances & 0 & 0 \\
      \hline
      Data collection, number of instances & 3 & 3 \\
      Data collection, total duration & 91.5s & 81.5s \\
      Data collection, non-overlapping instances & 0 & 0\\
      \hline
    \end{tabular}
    \caption{Details of interrater coding for a 10 minute sample of data.}
    \label{tab:Interrater}
\end{table}

\subsubsection{Phase 2}

\emph{Examples of interpretation of results and other discussions}

An example of speech coded as interpretation of results can be found on p. 13, as the final section of transcript in Sec \ref{SocialGroupAnalysis}. In that conversation, Group G discusses the meaning of their results from both experiments with the TA. 

Below is an example that was coded as other speech, despite the TA asking what the students found:

\begin{quote}
\textbf{TA:} What have you found\\
\textbf{Student 1:} Well it does that sometime and I don't think its supposed to\\
\textbf{TA:}  What is it doing?\\
\textbf{Student 1:}  Its like the velocity is going everywhere and like position\\
\textbf{TA:}  So have you not yet gotten a nice clean… This things in the way\\
\textbf{Student 1:} No like you see, it was before. It was doing it fine before, it didn’t matter\\
\textbf{TA:}  Did you ever get just like a nice clean bouncy bounce\\
\textbf{Student 1:}  Yeah \\
\textbf{Student 2:}  We got a couple that were nice\\
\end{quote}

As the primary substance of this interaction is on whether or not their data is ``clean" rather than on how to interpret their data.  

A subsection of 5 minutes of video was coded by two coders. Inter-rater agreement on what constituted ``Interpretation of Results" was 100\%, with some small variance (up to 5 seconds) for when to begin or end the code. This variability can be accounted for the two coders marking the transition between codes at slightly different times.

\subsection{Additional description of groups}
\label{SupplementGroups}
Here, we offer further description of the four groups that did problematize, Groups A-D. Descriptions also appear in \cite{ProblematizingPERC}. We also include timelines of all of the coding conducted in Phase 1 for all groups. Note that in the timeline figures, inaudible speech is coded as on- or off-topic where possible. For example, if one student offers an off topic contribution, another student responds inaudibly, and the first student continues the same off topic conversation apparently responding to the second student, the entire exchange was coded as off-topic. The percentages given in the bar graphs are taken as percentages of the entire episode, including full group discussions. Therefore, these percentages differ from those in Figs.\ref{fig:Audible}, \ref{fig:Topic} which coded only time in small groups.

\subsubsection{Group A}
\begin{figure*}[h]
    \centering
    \includegraphics[width=6in]{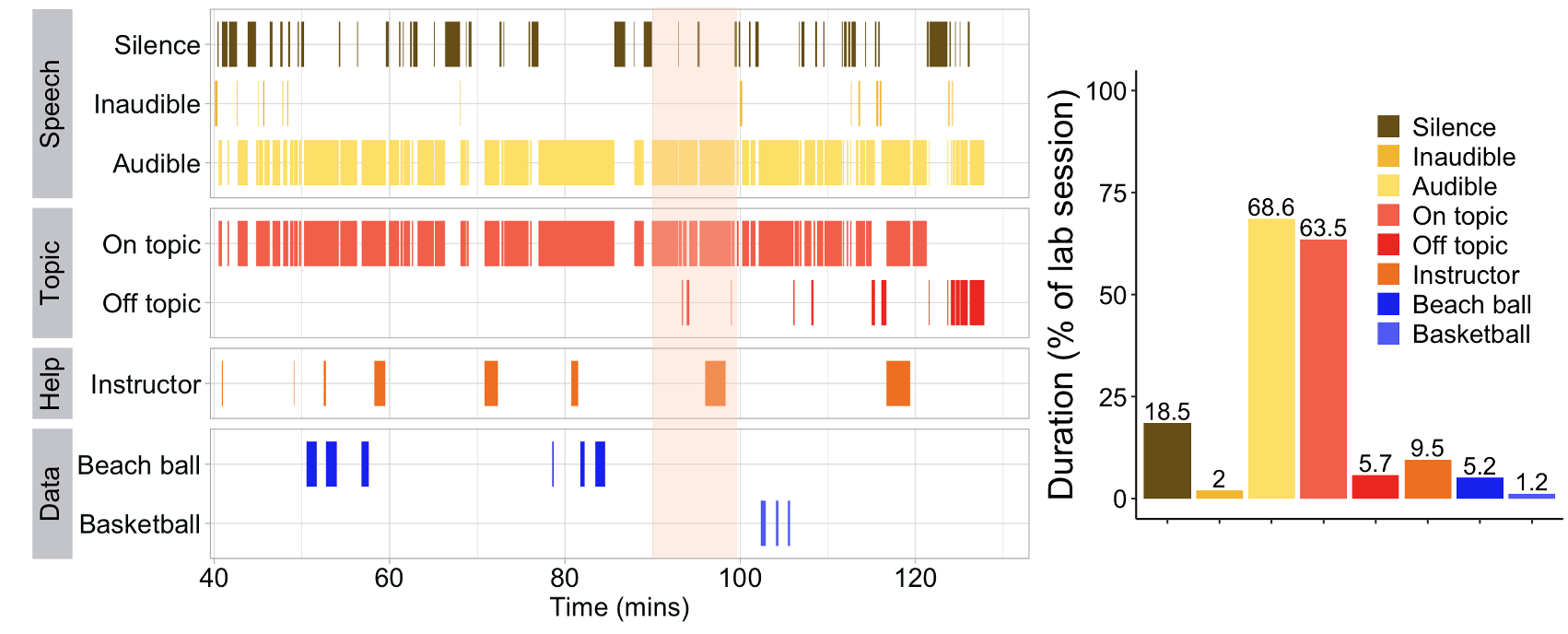}
    \caption{Group A full timeline highlighting episode analyzed in Ref. \cite{ProblematizingPERC} and durations of each behavior as a percentage of the video analyzed.}
    \label{fig:GroupA}
\end{figure*}

Group A consisted of one female and two male students. Throughout the lab, they often joke and take a playful approach to their investigations. When they encountered the designed problem of the lab, they jokingly posit that the moon is responsible for the extra upwards force. One of the male students often performs calculations and analysis and handles the lab equipment, while the female student often takes notes. The other male student alternates between all three tasks. He, along with the female student, are the primary participants in the joking and off-topic conversation. A timeline of how they were analyzed in Phase 1 is shown in Fig.\ref{fig:GroupA}

\subsubsection{Group B}
\begin{figure*}[h]
    \centering
    \includegraphics[width=6in]{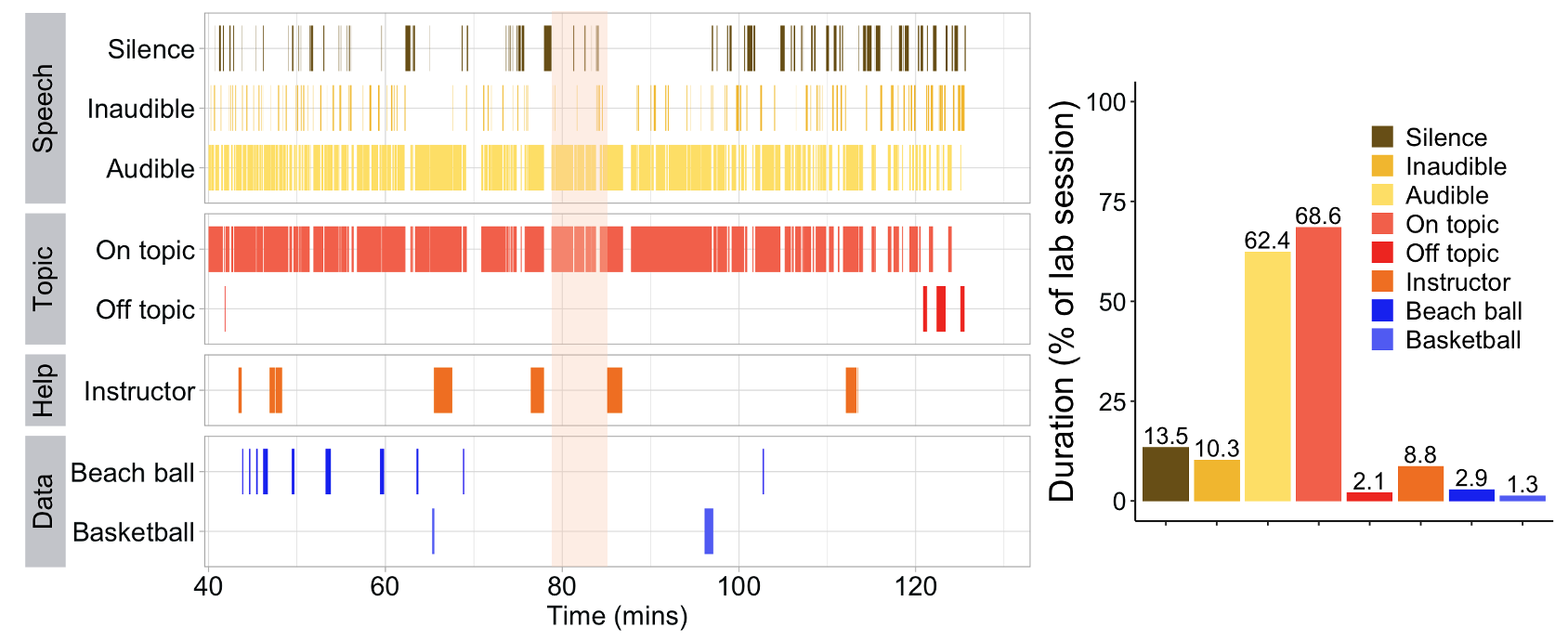}
    \caption{Group B full timeline highlighting episode analyzed in Ref. \cite{ProblematizingPERC} and durations of each behavior as a percentage of the video analyzed.}
    \label{fig:GroupB}
\end{figure*}

Group B consisted two female students and one male student. They multiple rounds of data at the start of the lab, running many trials until they are satisfied with their data. Their initial means of calculating acceleration differs from most other groups: they take an "initial" velocity near the start of the trajectory and a final velocity of zero at the top of the trajectory, subtract and divide by the time. At the prompting of the TA, they later switch to finding the slope of the velocity graph. They frequently laugh and smile, though their conversation is almost strictly on-topic. All three students frequently ask questions to their peers about what their data means, and participation among the three students is relatively balanced. A timeline of how they were analyzed in Phase 1 is shown in Fig.\ref{fig:GroupB}

\subsubsection{Group C}
\begin{figure*}[h]
    \centering
    \includegraphics[width=6in]{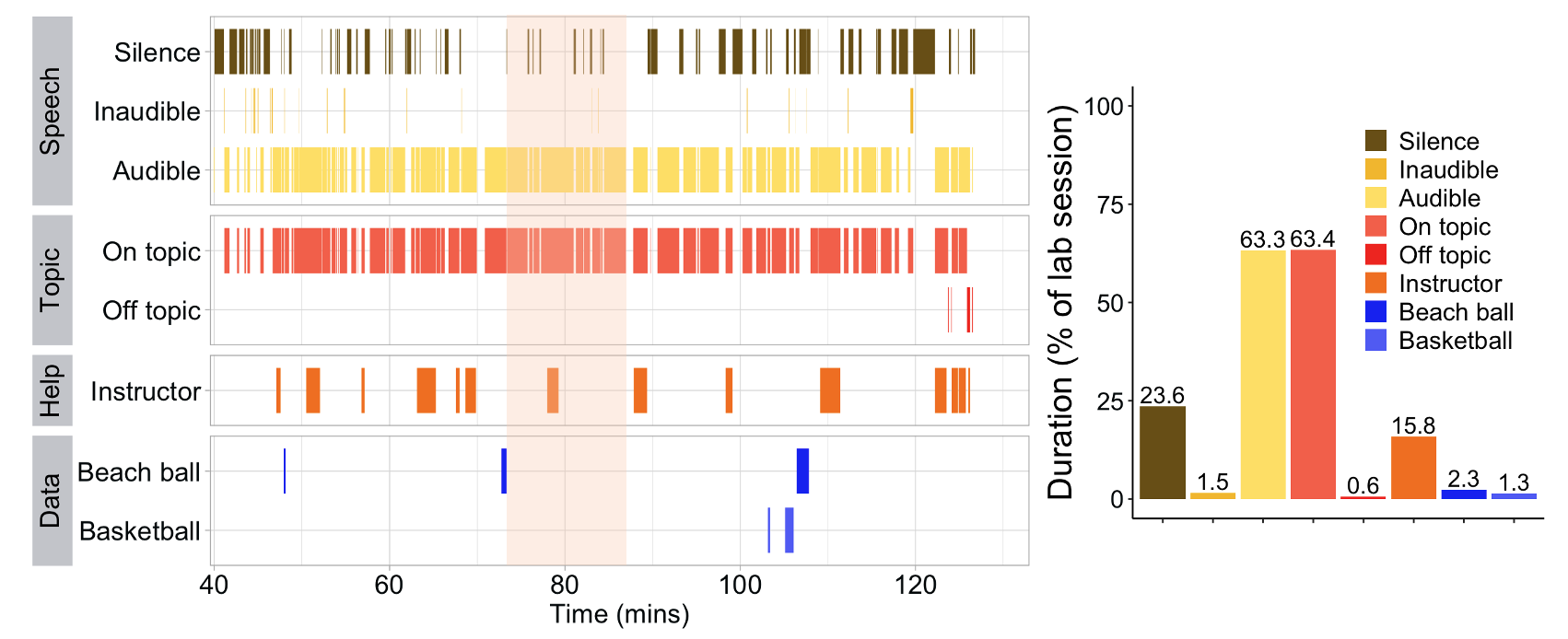}
    \caption{Group C full timeline highlighting episode analyzed in Ref. \cite{ProblematizingPERC} and durations of each behavior as a percentage of the video analyzed.}
    \label{fig:GroupC}
\end{figure*}

Group C consisted of one female and one male student. The male student speaks more than the female student, though her contributions often guide the conversation. She frequently highlights when something does or does not make sense to her. They are also quick to arrive at a possible explanation of ``airflow" causing the lower than expected acceleration and spend time reading about possible explanations on the internet. The female student conducts most of the hands on work and they alternating note-taking responsibilities. Near the end of the lab, they express that they have many questions, and their lab notes end with the statement ``I have to submit this now but I could easily type on this for at least another half hour this is very interesting," which appears to be written by the male student. A timeline of how they were analyzed in Phase 1 is shown in Fig.\ref{fig:GroupC}

\subsubsection{Group D}
\begin{figure*}[h]
    \centering
    \includegraphics[width=6in]{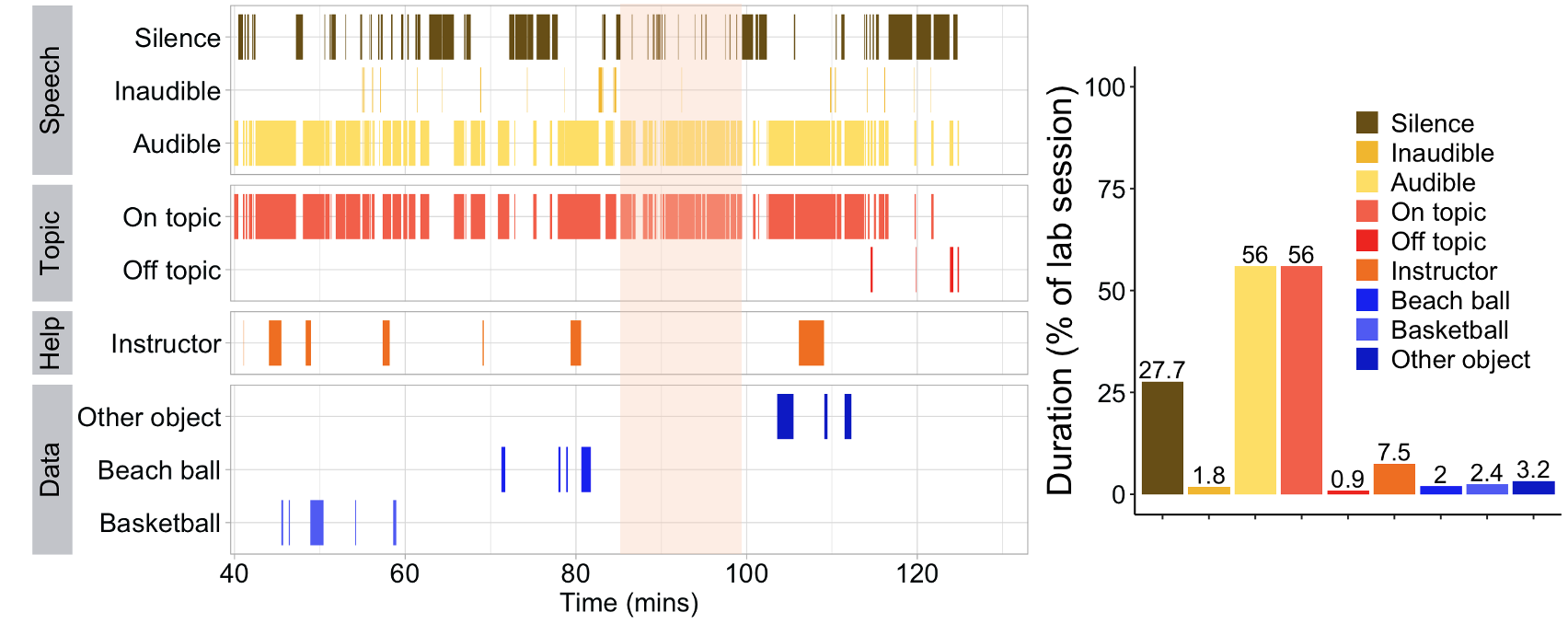}
    \caption{Group D full timeline highlighting episode analyzed in Ref. \cite{ProblematizingPERC} and durations of each behavior as a percentage of the video analyzed.}
    \label{fig:GroupD}
\end{figure*}

Group D consisted of three male students. Of all of the clearly audible groups, their conversation is the slowest paced and they spend the most time in silence. Two of the students dominate the conversation, though one of those two asks the third for his thoughts several times, indicating an awareness of this lack of balance in contributions. Almost immediately after acknowledging the designed problem in the lab, one student offers ``buoyancy" as possible explanation. However, they almost immediately abandon that explanation. Near the end of the lab, they decide to drop a number of different objects, including a empty soda bottle, a roll of tape, and a coffee filter to test out their idea that shape impacts an object's acceleration. They are the only group in this session to test objects other than the two balls. A timeline of how they were analyzed in Phase 1 is shown in Fig.\ref{fig:GroupD}

\subsubsection{Non-problematizing groups}
\begin{figure*}[h]
    \centering
    \includegraphics[width=6in]{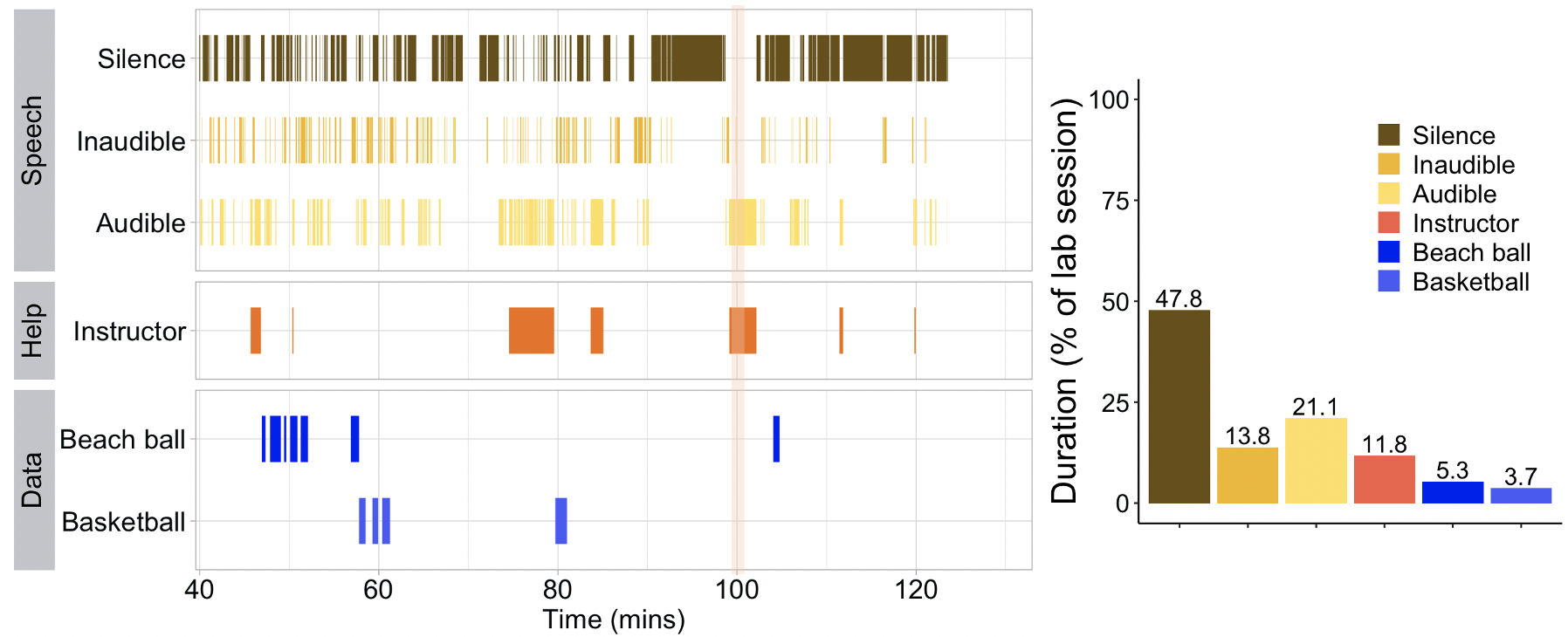}
    \caption{Group E full timeline  and durations of each behavior as a percentage of the video analyzed. Due to the extended silence and frequent inaudible speech, we were not able to identify on and off topic speech consistently. The portion analyzed in Sec.  \ref{AnalysisE} is highlighted.}
    \label{fig:GroupE}
\end{figure*}

\begin{figure*}[h]
    \centering
    \includegraphics[width=6in]{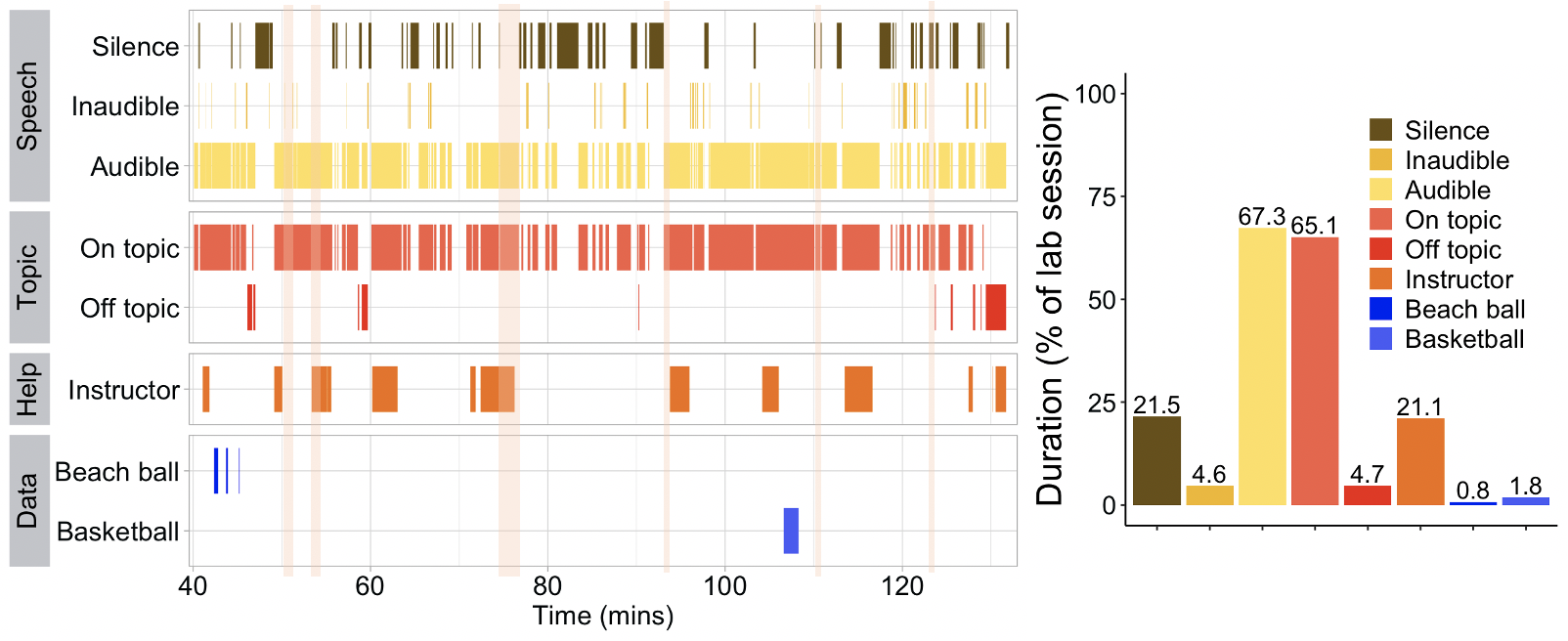}
    \caption{Group F full timeline  and durations of each behavior as a percentage of the video analyzed. The portions analyzed in Sec. \ref{AnalysisF} are highlighted.}
    \label{fig:GroupF}
\end{figure*}

\begin{figure*}[h]
    \centering
    \includegraphics[width=6in]{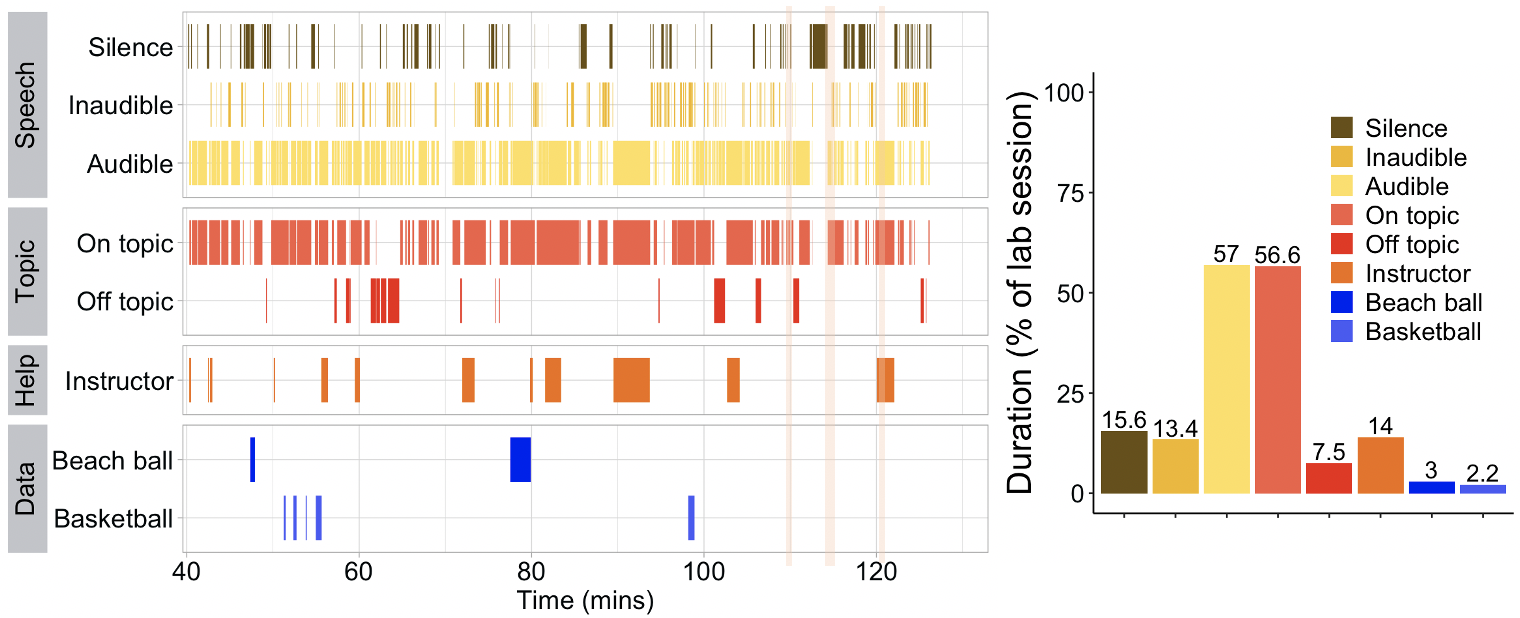}
    \caption{Group G full timeline and durations of each behavior as a percentage of the video analyzed. The portions analyzed in Sec. \ref{AnalysisG} are highlighted.}
    \label{fig:GroupG}
\end{figure*}

Timelines of how Groups, E, F, and G were analyzed in Phase 1 is shown in Figs. \ref{fig:GroupE},\ref{fig:GroupF}, and \ref{fig:GroupG}.

\end{document}